\documentclass[12pt]{iopart}
 
\usepackage{color}
\usepackage[mathenv]{}
\usepackage{graphicx}
\usepackage{natbib}
\usepackage{amssymb}

\usepackage{ulem}
\usepackage{color}

\begin{document}

\title[LIGO]{LIGO: The Laser Interferometer Gravitational-Wave Observatory}

\author{The LIGO Scientific Collaboration}
\author{B.~P.~Abbott$^{1}$,
R.~Abbott$^{1}$,
R.~Adhikari$^{1}$,
P.~Ajith$^{2}$,
B.~Allen$^{2,3}$,
G.~Allen$^{4}$,
R.~S.~Amin$^{5}$,
S.~B.~Anderson$^{1}$,
W.~G.~Anderson$^{3}$,
M.~A.~Arain$^{6}$
% ...
\\
(The LIGO Scientific Collaboration, http://www.ligo.org)}


\address{$^{1}$LIGO - California Institute of Technology, Pasadena, CA  91125, USA}
\address{$^{2}$Albert-Einstein-Institut, Max-Planck-Institut f\"ur Gravitationsphysik, D-30167 Hannover, Germany}
\address{$^{3}$University of Wisconsin-Milwaukee, Milwaukee, WI 53201, USA}
\address{$^{4}$Stanford University, Stanford, CA 94305, USA}
\address{$^{5}$Louisiana State University, Baton Rouge, LA 70803, USA}
\address{$^{6}$University of Florida, Gainesville, FL  32611, USA}
% ...


\ead{peter.fritschel@ligo.org}

\date{}

\definecolor{purple}{rgb}{0.6,0,1}
\pagestyle{myheadings}
\markboth{\color{purple}{LIGO-P070082-06}}{\color{purple}{LIGO-P070082-06}}



\begin{abstract}
The goal of the Laser Interferometric
Gravitational-Wave Observatory (LIGO) is to detect and study
gravitational waves of astrophysical origin. Direct detection of
gravitational waves holds the promise of testing general relativity
in the strong-field regime, of providing a new probe of exotic
objects such as black holes and neutron stars, and of uncovering
unanticipated new astrophysics. LIGO, a joint Caltech-MIT project
supported by the National Science Foundation, operates three
multi-kilometer interferometers at two widely separated sites in the
United States. These detectors are the result of decades of
worldwide technology development, design, construction, and
commissioning. They are now operating at their design sensitivity,
and are sensitive to gravitational wave strains smaller than one part
in $10^{21}$. With this unprecedented sensitivity, the data are
being analyzed to detect or place limits on gravitational waves from 
a variety of potential astrophysical sources.
\end{abstract}

\submitto{\RPP}
\maketitle

\section{Introduction}
The prediction of gravitational waves (GWs), oscillations in the
space-time metric that propagate at the speed of light, is one of
the most profound differences between Einstein's general theory of
relativity and the Newtonian theory of gravity that it replaced. GWs
remained a theoretical prediction for more than 50 years until the
first observational evidence for their existence came with the
discovery and subsequent observations of the binary pulsar PSR 1913+16, 
by Russell Hulse and Joseph Taylor. This is a system of two neutron stars 
that orbit each other with a period of 7.75 hours. Precise timing of radio 
pulses emitted by one of the neutron stars, monitored now over several
decades, shows that their orbital period is slowly decreasing at just 
the rate predicted for the general-relativistic emission of GWs
\cite{2005ASPC..328...25W}. Hulse and Taylor were awarded the Nobel Prize in
Physics for this work in 1993.

In about 300 million years, the PSR 1913+16 orbit will decrease to the
point where the pair coalesces into a single compact object, a process
that will produce directly detectable gravitational waves. In the
meantime, the direct detection of GWs will require similarly strong
sources\,--\,extremely large masses moving with large accelerations in
strong gravitational fields. The goal of LIGO, the Laser
Interferometer Gravitational-Wave Observatory \cite{Abramovici:1992ah}
is just that: to detect and study GWs of astrophysical origin.
Achieving this goal will mark the opening of a new window on the
universe, with the promise of new physics and astrophysics. In
physics, GW detection could provide information about strong-field
gravitation, the untested domain of strongly curved space where
Newtonian gravitation is no longer even a poor approximation. In
astrophysics, the sources of GWs that LIGO might detect include binary
neutron stars (like PSR 1913+16 but much later in their evolution);
binary systems where a black hole replaces one or both of the neutron
stars; a stellar core collapse which triggers a Type II supernova;
rapidly rotating, non-axisymmetric neutron stars; and possibly
processes in the early universe that produce a stochastic background
of GWs \cite{2002gr.qc.....4090C}.

In the past few years the field has reached a milestone, with
decades-old plans to build and operate large interferometric GW
detectors now realized in several locations worldwide. This article
focuses on LIGO, which operates the most sensitive detectors yet
built. We aim to describe the LIGO detectors and how they operate,
explain how they have achieved their remarkable sensitivity, and
review how their data can be used to learn about a variety of
astrophysical phenomena.

\section{Gravitational waves}

The essence of general relativity is that mass and energy produce a
curvature of four-dimensional space-time, and that matter moves in
response to this curvature. The Einstein field equations prescribe the
interaction between mass and space-time curvature, much as Maxwell's
equations prescribe the relationship between electric charge and
electromagnetic fields. Just as electromagnetic waves are
time-dependent vacuum solutions to Maxwell's equations, gravitational
waves are time-dependent vacuum solutions to the field
equations. Gravitational waves are oscillating perturbations to a
flat, or Minkowski, space-time metric, and can be thought of
equivalently as an oscillating strain in space-time or as an
oscillating tidal force between free test masses.

\begin{figure}
\begin{center}
\includegraphics[width=12cm]{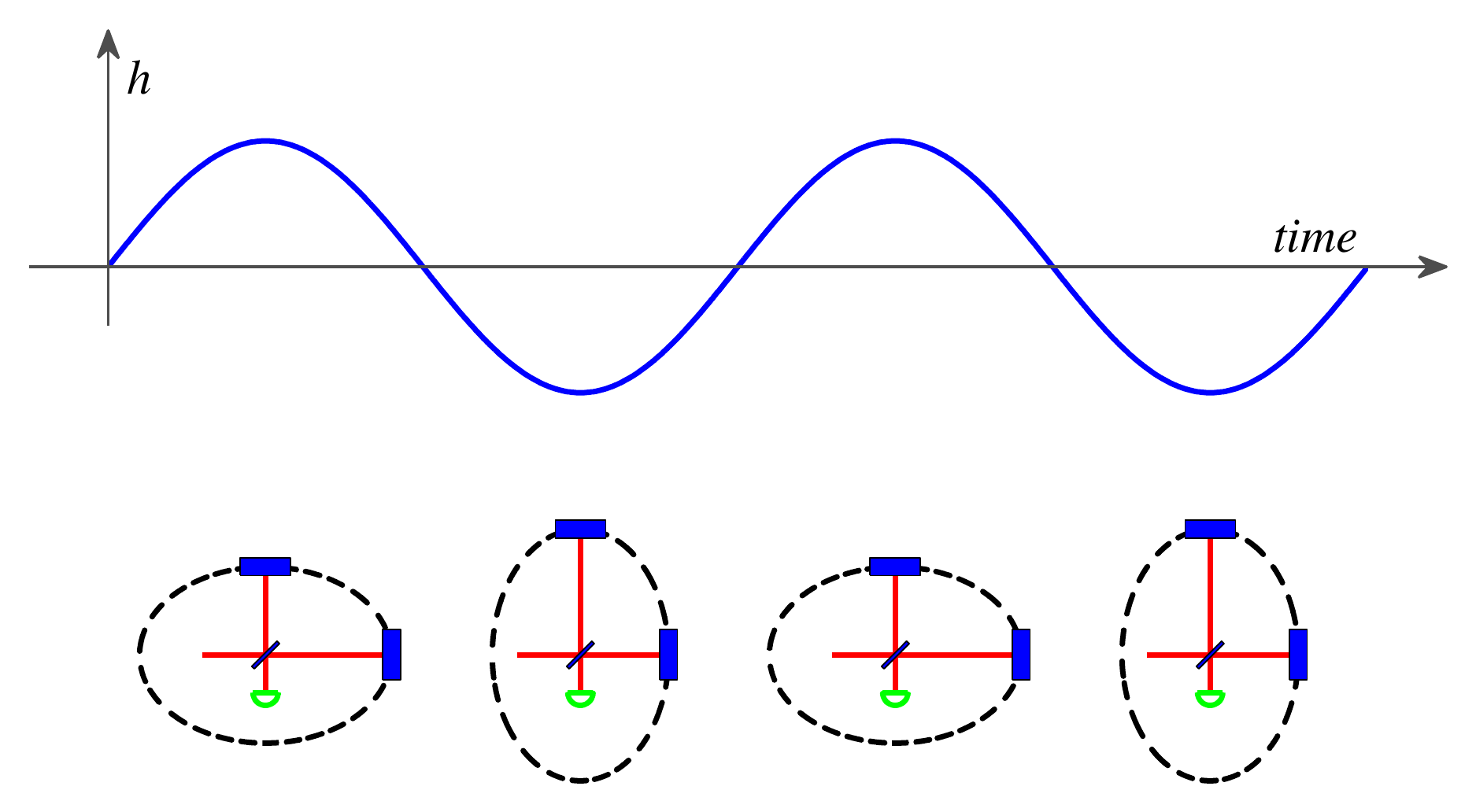}
\caption{A gravitational wave traveling perpendicular to the plane of
  the diagram is characterized by a strain amplitude $h$. The wave
  distorts a ring of test particles into an ellipse, elongated in one
  direction in one half-cycle of the wave, and elongated in the
  orthogonal direction in the next half-cycle. This oscillating 
  distortion can be measured with a Michelson interferometer oriented
  as shown. The length oscillations modulate the phase shifts accrued by
  the light in each arm, which are in turn observed as light intensity
  modulations at the photodetector (green semi-circle). This depicts one
  of the linear polarization modes of the GW. \label{fig:GWcartoon}}
\end{center}
\end{figure}

As with electromagnetic waves, gravitational waves travel at the speed
of light and are transverse in character\,--\,\textit{i.e.}, the
strain oscillations occur in directions orthogonal to the direction
the wave is propagating. Whereas electromagnetic waves are dipolar in
nature, gravitational waves are quadrupolar: the strain pattern
contracts space along one transverse dimension, while expanding it
along the orthogonal direction in the transverse plane (see
Fig.~\ref{fig:GWcartoon}). Gravitational radiation is produced by
oscillating multipole moments of the mass distribution of a
system. The principle of mass conservation rules out monopole
radiation, and the principles of linear and angular momentum
conservation rule out gravitational dipole radiation. Quadrupole
radiation is the lowest allowed form, and is thus usually the dominant
form. In this case, the gravitational wave field strength is
proportional to the second time derivative of the quadrupole moment of
the source, and it falls off in amplitude inversely with distance from
the source. The tensor character of gravity\,--\,the hypothetical
graviton is a spin-2 particle\,--\,means that the transverse strain
field comes in two orthogonal polarizations. These are commonly
expressed in a linear polarization basis as the `$+$' polarization
(depicted in Fig.~\ref{fig:GWcartoon}) and the `$\times$'
polarization, reflecting the fact that they are rotated 45 degrees
relative to one another. An astrophysical GW will, in general, be a
mixture of both polarizations.

Gravitational waves differ from electromagnetic waves in that they
propagate essentially unperturbed through space, as they interact only
very weakly with matter. Furthermore, gravitational waves are
intrinsically non-linear, because the wave energy density itself
generates additional curvature of space-time. This phenomenon is only
significant, however, very close to strong sources of waves, where the
wave amplitude is relatively large. More usually, gravitational waves
distinguish themselves from electromagnetic waves by the fact that
they are very weak. One cannot hope to detect any waves of terrestrial
origin, whether naturally occurring or manmade; instead one must look
to very massive compact astrophysical objects, moving at relativistic
velocities. For example, strong sources of gravitational waves that
may exist in our galaxy or nearby galaxies are expected to produce
wave strengths on Earth that do not exceed strain levels of one part
in $10^{21}$. Finally, it is important to appreciate that GW detectors
respond directly to GW amplitude rather than GW power; therefore the
volume of space that is probed for potential sources increases as the
cube of the strain sensitivity.

\section{LIGO and the worldwide detector network}

\begin{figure}
\begin{center}
\includegraphics[width=11cm]{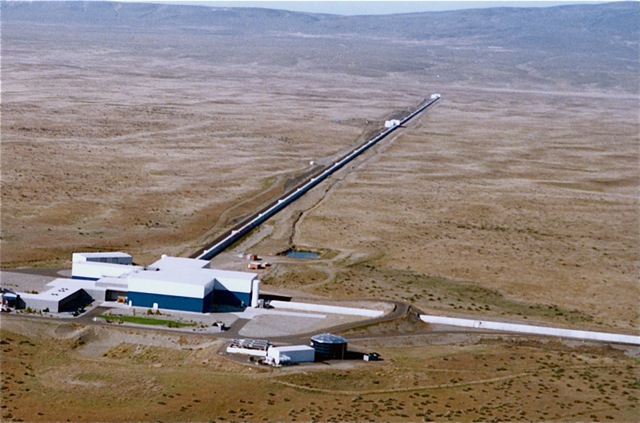}
\includegraphics[width=11cm]{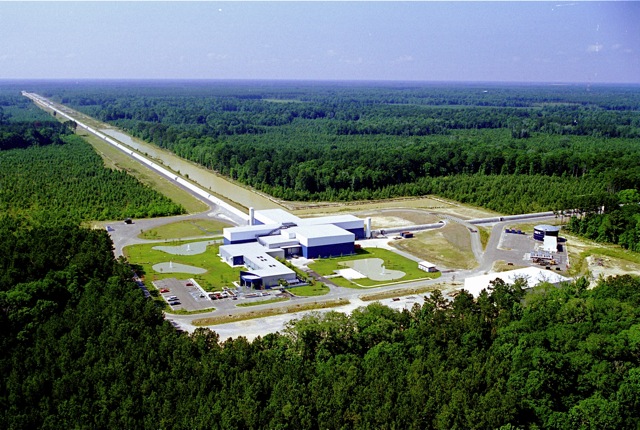}
\caption{Aerial photograph of the LIGO observatories at Hanford,
  Washington (top) and Livingston, Louisiana (bottom). The lasers 
  and optics are contained in the white and blue buildings. From 
  the large corner building, evacuated beam tubes extend at right
  angles for 4 km in each direction (the full length of only one
  of the arms is seen in each photo); the tubes are covered by the 
  arched, concrete enclosures seen here. 
\label{fig:observatories}}
\end{center}
\end{figure}

As illustrated in Fig.~\ref{fig:GWcartoon}, the oscillating
quadrupolar strain pattern of a GW is well matched by a Michelson
interferometer, which makes a very sensitive comparison of the lengths
of its two orthogonal arms. LIGO utilizes three specialized Michelson
interferometers, located at two sites (see
Fig.~\ref{fig:observatories}): an observatory on the Hanford site in
Washington houses two interferometers, the 4 km-long H1 and 2 km-long
H2 detectors; and an observatory in Livingston Parish, Louisiana,
houses the 4 km-long L1 detector. Other than the shorter length of H2,
the three interferometers are essentially identical. Multiple
detectors at separated sites are crucial for rejecting instrumental
and environmental artifacts in the data, by requiring coincident
detections in the analysis. Also, because the antenna pattern of an
interferometer is quite wide, source localization requires
triangulation using three separated detectors.

The initial LIGO detectors were designed to be sensitive to GWs in the
frequency band 40\,--\,7000\,Hz, and capable of detecting a GW strain
amplitude as small as $10^{-21}$ \cite{Abramovici:1992ah}. With
funding from the National Science Foundation, the LIGO sites and
detectors were designed by scientists and engineers from the
California Institute of Technology and the Massachusetts Institute of
Technology, constructed in the late 1990s, and commissioned over the
first 5 years of this decade. From November 2005 through September
2007, they operated at their design sensitivity in a continuous
data-taking mode. The data from this science run, known as S5, are
being analyzed for a variety of GW signals by a group of researchers
known as the LIGO Scientific Collaboration \cite{LSChomepage}.  At the
most sensitive frequencies, the instrument root-mean-square (rms)
strain noise has reached an unprecedented level of $3\times10^{-22}$
in a 100 Hz band.

Although in principle LIGO can detect and study GWs by itself, the
potential to do astrophysics can be quantitatively and qualitatively
enhanced by operation in a more extensive network. For example, the
direction of travel of the GWs and the complete polarization
information carried by the waves can only be extracted by a network of
detectors. Such a global network of GW observatories has been emerging
over the past decade. In this period, the Japanese TAMA project built
a 300~m interferometer outside Tokyo, Japan \cite{Takahashi:2004rwa}; the
German-British GEO project built a 600~m interferometer near Hanover,
Germany \cite{Luck:2006ug}; and the European Gravitational Observatory
built the 3 km-long interferometer Virgo near Pisa, Italy
\cite{Acernese:2006bj}. In addition, plans are underway to develop a
large scale gravitational wave detector in Japan sometime during the
next decade \cite{0264-9381-23-8-S27}.

Early in its operation LIGO joined with the GEO project; for strong
sources the shorter, less sensitive GEO\,600 detector provides added
confidence and directional and polarization information.  In May 2007
the Virgo detector began joint observations with LIGO, with a strain
sensitivity close to that of LIGO's 4~km interferometers at
frequencies above $\sim\!1$~kHz. The LIGO Scientific Collaboration and
the Virgo Collaboration negotiated an agreement that all data
collected from that date are to be analyzed and published jointly.

\section{Detector description} 

Figure~\ref{fig:GWcartoon} illustrates the basic concept of how a
Michelson interferometer is used to measure a GW strain. The challenge
is to make the instrument sufficiently sensitive: at the targeted
strain sensitivity of $10^{-21}$, the resulting arm length change is
only $\sim\!10^{-18}$~m, a thousand times smaller than the diameter of
a proton. Meeting this challenge involves the use of special
interferometry techniques, state-of-the-art optics, highly stable
lasers, and multiple layers of vibration isolation, all of which are
described in the sections that follow. And of course a key feature of
the detectors is simply their scale: the arms are made as long as
practically possible to increase the signal due to a GW strain.  See
Table~\ref{tab:params} for a list of the main design parameters of the
LIGO interferometers.

\subsection{Interferometer Configuration}

The LIGO detectors are Michelson interferometers whose mirrors also
serve as gravitational test masses. A passing gravitational wave will
impress a phase modulation on the light in each arm of the Michelson,
with a relative phase shift of 180 degrees between the arms. When the
Michelson arm lengths are set such that the un-modulated light
interferes destructively at the antisymmetric (AS) port\,--\,the dark
fringe condition\,--\,the phase modulated sideband light will
interfere constructively, with an amplitude proportional to GW strain
and the input power. With dark fringe operation, the full power incident on
the beamsplitter is returned to the laser at the symmetric port.  Only
differential motion of the arms appears at the AS port; common
mode signals are returned to the laser with the carrier light.

\begin{figure}[t]
\begin{center}
\includegraphics[width=14cm]{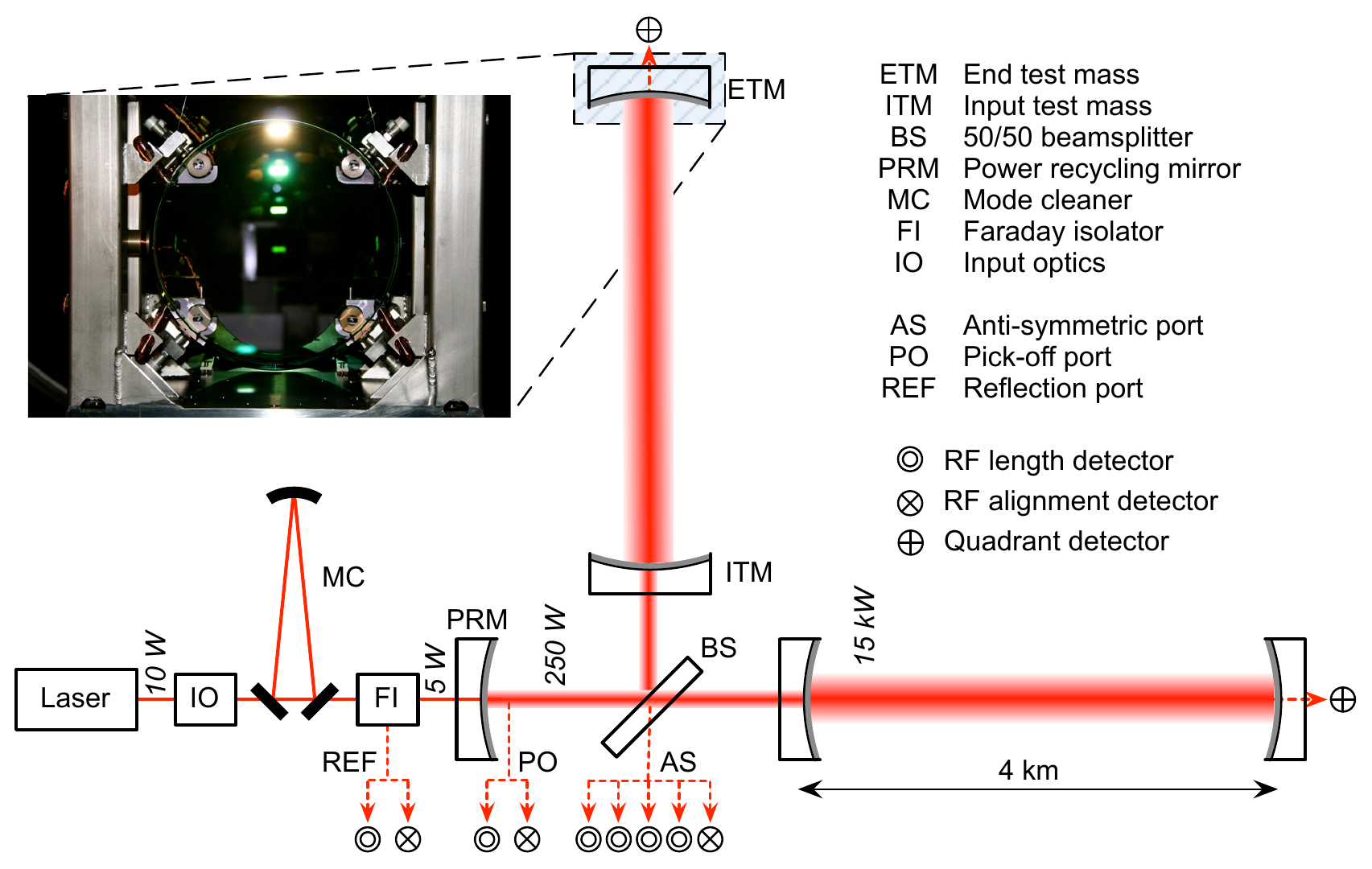}
\caption{Optical and sensing configuration of the LIGO 4~km
  interferometers (the laser power numbers here are generic; specific
  power levels are given in Table~1). The IO block includes laser
  frequency and amplitude stabilization, and electro-optic phase
  modulators.  The power recycling cavity is formed between the PRM
  and the two ITMs, and contains the BS. The inset photo shows an input
  test mass mirror in its pendulum suspension. The near face has a
  highly reflective coating for the infrared laser light, but transmits
  visible light. Through it one can see mirror actuators
  arranged in a square pattern near the mirror perimeter.\label{fig:schematic}}
\end{center}
\end{figure}

Two modifications to a basic Michelson, shown in
Fig.~\ref{fig:schematic}, increase the carrier power in the arms and
hence the GW sensitivity. First, each arm contains a resonant
Fabry-Perot optical cavity made up of a partially transmitting input
mirror and a high reflecting end mirror.  The cavities cause the light
to effectively bounce back and forth multiple times in the arms,
increasing the carrier power and phase shift for a given strain
amplitude. In the LIGO detectors the Fabry-Perot cavities multiply the
signal by a factor of 100 for a 100~Hz GW.  Second, a
partially-reflecting mirror is placed between the laser and
beamsplitter to implement power recycling \cite{Meers:1988wp}. In this
technique, an optical cavity is formed between the power recycling
mirror and the Michelson symmetric port. By matching the transmission
of the recycling mirror to the optical losses in the Michelson, and
resonating this recycling cavity, the laser power stored in the
interferometer can be significantly increased. In this configuration,
known as a power recycled Fabry-Perot Michelson, the LIGO
interferometers increase the power in the arms by a factor of $\approx
8,000$ with respect to a simple Michelson.


\subsection{Laser and Optics}
\label{sec:lasers}
The laser source is a diode-pumped, Nd:YAG master oscillator and power
amplifier system, and emits 10~W in a single frequency at 1064~nm
\cite{Savage:1998}. The laser power and frequency are actively
stabilized, and passively filtered with a transmissive ring cavity
(pre-mode cleaner, PMC). The laser power stabilization is implemented
by directing a sample of the beam to a photodetector, filtering its
signal and feeding it back to the power amplifier; this servo
stabilizes the relative power fluctuations of the beam to
$\sim\!10^{-7} /\sqrt{\rm Hz}$ at 100~Hz \cite{abbott:1346}.  The
laser frequency stabilization is done in multiple stages that are more
fully described in later sections. The first, or pre-stabilization
stage uses the traditional technique of servo locking the laser
frequency to an isolated reference cavity using the Pound-Drever-Hall
(PDH) technique \cite{1983ApPhB..31...97D}, in this case via feedback
to frequency actuators on the master oscillator and to an
electro-optic phase modulator. The servo bandwith is 500~kHz, and the
pre-stabilization achieves a stability level of $\sim\!10^{-2}\,{\rm
  Hz/\sqrt{Hz}}$ at 100~Hz.  The PMC transmits the pre-stabilized
beam, filtering out both any light not in the fundamental Gaussian
spatial mode and laser noise at frequencies above a few MHz
\cite{Willke:98}. The PMC output beam is weakly phase-modulated with
two radio-frequency (RF) sine waves, producing, to first-order, two
pairs of sideband fields around the carrier field; these RF sideband
fields are used in a heterodyne detection system described below.

After phase modulation, the beam passes into the LIGO vacuum system.
All the main interferometer optical components and beam paths are
enclosed in the ultra-high vacuum system
($10^{-8}$\,--\,$10^{-9}$~torr) for acoustical isolation and to reduce
phase fluctuations from light scattering off residual gas
\cite{1996magr.meet.1434Z}. The long beam tubes are particularly
noteworthy components of the LIGO vacuum system. These 1.2~m diameter,
4~km long stainless steel tubes were designed to have low-outgassing
so that the required vacuum could be attained by pumping only from the
ends of the tubes. This was achieved by special processing of the
steel to remove hydrogen, followed by an \textit{in-situ} bakeout of
the spiral-welded tubes, for approximately 20 days at 160~C.

The in-vacuum beam first passes through the mode cleaner (MC), a 12~m
long, vibrationally isolated transmissive ring cavity. The MC provides
a stable, diffraction-limited beam with additional filtering of laser
noise above several kilohertz \cite{skeldon:2443}, and it serves as an
intermediate reference for frequency stabilization. The MC length and
modulation frequencies are matched so that the main carrier field and
the modulation sideband fields all pass through the MC. After the MC
is a Faraday isolator and a reflective 3-mirror telescope that expands
the beam and matches it to the arm cavity mode.

The interferometer optics, including the test masses, are fused-silica
substrates with multilayer dielectric coatings, manufactured to have
extremely low scatter and low absorption.  The test mass substrates
are polished so that the surface deviation from a spherical figure,
over the central 80~mm diameter, is typically 5 angstroms or smaller,
and the surface microroughness is typically less than 2 angstroms
\cite{Walsh:99}. The mirror coatings are made using ion-beam
sputtering, a technique known for producing ultralow-loss mirrors
\cite{Wei:89, Rempe:92}. The absorption level in the coatings
is generally a few parts-per-million (ppm) or less \cite{Ottaway:06},
and the total scattering loss from a mirror surface is estimated to be
60\,--\,70~ppm.

In addition to being a source of optical loss, scattered light can be
a problematic noise source, if it is allowed to reflect or scatter
from a vibrating surface (such as a vacuum system wall) and recombine
with the main beam \cite{Vinet:1996uf}. Since the vibrating,
re-scattering surface may be moving by $\sim\!10$ orders of magnitude
more than the test masses, very small levels of scattered light can
contaminate the output. To control this, various baffles are employed
within the vacuum system to trap scattered light \cite{Vinet:1996uf,
  Vinet:1997jx}. Each 4~km long beam tube contains approximately two
hundred baffles to trap light scattered at small angles from the test
masses. These baffles are stainless steel truncated cones, with
serrated inner edges, distributed so as to completely hide the beam
tube from the line of sight of any arm cavity mirror. Additional
baffles within the vacuum chambers prevent light outside the mirror
apertures from hitting the vacuum chamber walls.

\renewcommand{\arraystretch}{1.3}
\begin{table}
\begin{center}
\begin{tabular}{lccc}
   &  {\bf H1}  &  {\bf L1}  &  {\bf H2} \\
\hline\hline
Laser type and wavelength & \multicolumn{3}{c}{Nd:YAG, $\lambda = 1064$~nm} \\
Arm cavity finesse & \multicolumn{3}{c}{220} \\
Arm length & 3995 m & 3995 m & 2009 m \\
Arm cavity storage time, $\tau_s$ & 0.95 ms &  0.95 ms & 0.475 ms \\
Input power at recycling mirror & 4.5 W & 4.5 W & 2.0 W \\
Power Recycling gain  & 60 & 45 & 70 \\
Arm cavity stored power  & 20 kW & 15 kW & 10 kW \\
Test mass size \& mass  &  \multicolumn{3}{c}{$\phi\,25\,{\rm cm}\times10\,{\rm cm}$, 
   10.7 kg} \\
Beam radius ($1/e^2$ power) ITM/ETM & 3.6\,cm\,/\,4.5\,cm & 3.9\,cm\,/\,4.5\,cm &
  3.3\,cm\,/\,3.5\,cm \\
Test mass pendulum frequency & \multicolumn{3}{c}{0.76 Hz} \\
\hline\hline
\end{tabular}
\caption{Parameters of the LIGO interferometers. H1 and H2 refer to
the interferometers at Hanford, Washington, and L1 is the
interferometer at Livingston Parish, Louisiana.} \label{tab:params}
\end{center}
\end{table}
\renewcommand{\arraystretch}{1}

\subsection{Suspensions and Vibration Isolation}
Starting with the MC, each mirror in the beam line is suspended as a
pendulum by a loop of steel wire. The pendulum provides $f^{-2}$
vibration isolation above its eigenfrequencies, allowing free movement
of a test mass in the GW frequency band. Along the beam direction, a
test mass pendulum isolates by a factor of nearly $2 \times 10^{4}$ at
100~Hz. The position and orientation of a suspended optic is
controlled by electromagnetic actuators: small magnets are bonded to
the optic and coils are mounted to the suspension support structure,
positioned to maximize the magnetic force and minimize ground noise
coupling. The actuator assemblies also contain optical sensors
that measure the position of the suspended optic with respect to
its support structure. These signals are used to actively damp
eigenmodes of the suspension.

The bulk of the vibration isolation in the GW band is provided by
four-layer mass-spring isolation stacks, to which the pendulums are
mounted. These stacks provide approximately $f^{-8}$ isolation above
$\sim\!10$~Hz \cite{giaime:208}, giving an isolation factor of about
$10^8$ at 100~Hz. In addition, the L1 detector, subject to higher
environmental ground motion than the Hanford detectors, employs
seismic pre-isolators between the ground and the isolation
stacks. These active isolators employ a collection of motion sensors,
hydraulic actuators, and servo controls; the pre-isolators actively
suppress vibrations in the band $0.1-10$~Hz, by as much as a factor of
10 in the middle of the band \cite{Abbott:2004uh}.

\subsection{Sensing and Controls}

The two Fabry-Perot arms and power recycling cavities are essential to
achieving the LIGO sensitivity goal, but they require an active
feedback system to maintain the interferometer at the proper operating
point \cite{Fritschel:01}.  The round trip length of
each cavity must be held to an integer multiple of the laser
wavelength so that newly introduced carrier light interferes
constructively with light from previous round trips.  Under these
conditions the light inside the cavities builds up and they are said
to be on resonance.  In addition to the three cavity lengths, the
Michelson phase must be controlled to ensure that the AS port remains on
the dark fringe.

The four lengths are sensed with a variation of the PDH reflection
scheme \cite{Regehr:95}.  In standard PDH, an error signal is
generated through heterodyne detection of the light reflected from a
cavity. The RF phase modulation sidebands are directly reflected from
the cavity input mirror and serve as a local oscillator to mix with
the carrier field. The carrier experiences a phase-shift in
reflection, turning the RF phase modulation into RF amplitude
modulation, linear in amplitude for small deviations from
resonance. This concept is extended to the full interferometer as
follows. At the operating point, the carrier light is resonant in the
arm and recycling cavities and on a Michelson dark fringe.  The RF
sideband fields resonate differently.  One pair of RF sidebands (from
phase modulation at 62.5~MHz) is not resonant and simply reflects from
the recycling mirror. The other pair (25~MHz phase modulation) is
resonant in the recycling cavity but not in the arm
cavities.\footnote{These are approximate modulation frequencies for H1
  and L1; those for H2 are about 10\% higher.} The Michelson mirrors
are positioned to make one arm 30~cm longer than the other so that
these RF sidebands are not on a Michelson dark fringe. By design this
Michelson asymmetry is chosen so that most of the resonating RF
sideband power is coupled to the AS port.

In this configuration, heterodyne error signals for the four
length degrees-of-freedom are extracted from the three output
ports shown in Fig.~\ref{fig:schematic} (REF, PO and AS ports).
The AS port is heterodyned at the resonating RF frequency and 
gives an error signal proportional to differential arm length
changes, including those due to a GW. The PO port is a sample
of the recycling cavity beam, and is detected at the
resonating RF frequency to give error signals for the recycling
cavity length and the Michelson phase (using both RF quadratures).
The REF port is detected at the non-resonating RF frequency and
gives a standard PDH signal proportional to deviations in the laser frequency
relative to the average length of the two arms.

Feedback controls derived from these errors signals are applied to the
two end mirrors to stabilize the differential arm length, to
the beamsplitter to control the Michelson phase, and to the recycling mirror
to control the recycling cavity length. The feedback signals are applied
directly to the mirrors through their coil-magnet actuators, with slow
corrections for the differential arm length applied with longer-range
actuators that move the whole isolation stack.

\begin{figure}[t]
  \centering
  \includegraphics[width=14cm]{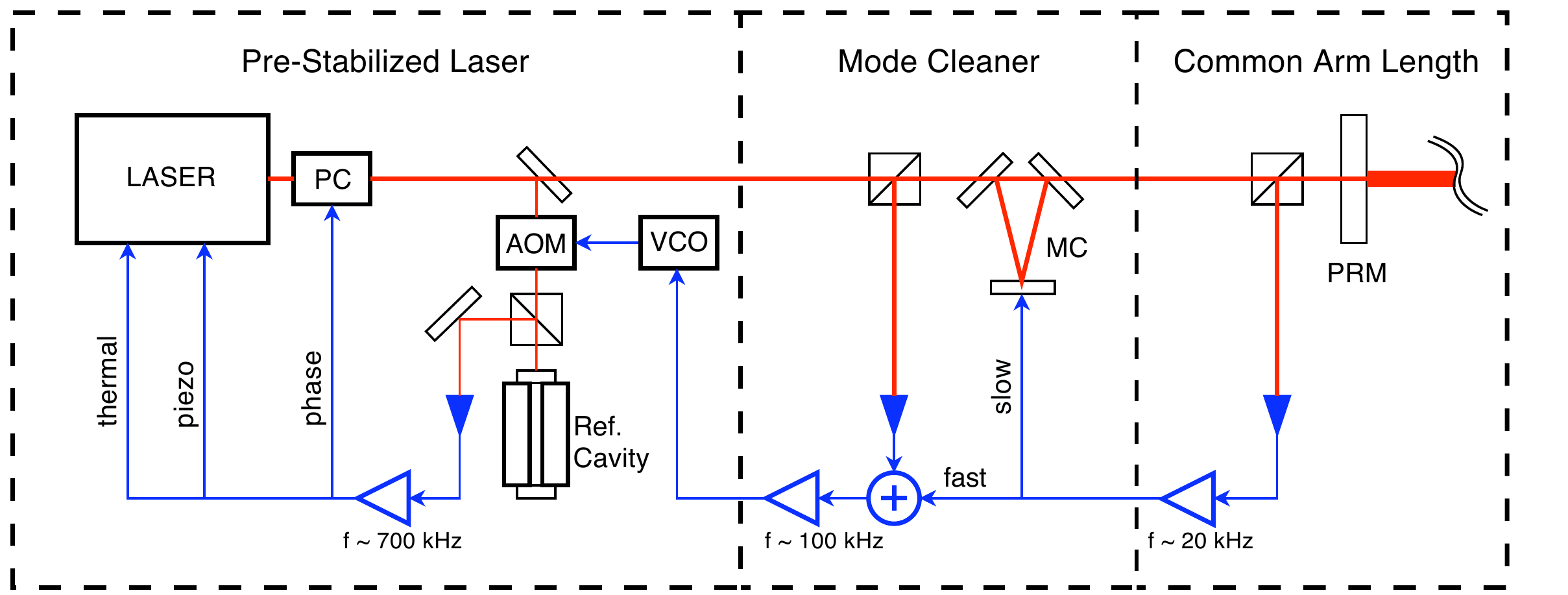}
  \caption{Schematic layout of the frequency stabilization servo.  The
    laser is locked to a fixed-length reference cavity through an AOM.
    The AOM frequency is generated by a Voltage Controlled Oscillator
    (VCO) driven by the MC, which is in turn driven by the common mode
    arm length signal from the REF port. The laser frequency is
    actuated by a combination of a Pockels Cell (PC), piezo actuator,
    and thermal control. }
  \label{fig:common_mode} 
\end{figure}

The common arm length signal from the REF port detection is used in
the final level of laser frequency stabilization \cite{Rana2004}
pictured schematically in Fig.~\ref{fig:common_mode}. The
hierarchical frequency control starts with the reference cavity
pre-stabilization mentioned in Sec.~\ref{sec:lasers}.  The
pre-stabilization path includes an Acousto-Optic Modulator (AOM)driven by a
voltage-controlled oscillator, through which fast corrections to the
pre-stabilized frequency can be made. The MC servo uses this
correction path to stabilize the laser frequency to the MC length,
with a servo bandwidth close to 100~kHz.  The most stable frequency
reference in the GW band is naturally the average length of the two
arm cavities, therefore the common arm length error signal provides
the final level of frequency correction. This is accomplished with
feedback to the MC, directly to the MC length at low frequencies and
to the error point of the MC servo at high frequencies, with an
overall bandwidth of 20~kHz. The MC servo then impresses the
corrections onto the laser frequency. The three cascaded frequency
loops\,--\,the reference cavity pre-stabilization; the MC loop; and
the common arm length loop\,--\,together provide 160~dB of frequency
noise reduction at 100~Hz, and achieve a frequency stability of
$5\,\mu$Hz rms in a 100~Hz bandwidth.

The photodetectors are all located outside the vacuum system, mounted
on optical tables. Telescopes inside the vacuum reduce the beam size
by a factor of $\sim\!10$, and the small beams exit the vacuum through
high-quality windows. To reduce noise from scattered light and
beam clipping, the optical tables are housed in acoustical enclosures,
and the more critical tables are mounted on passive vibration isolators.
Any back-scattered light along the AS port path is further mitigated with
a Faraday isolator mounted in the vacuum system.

The total AS port power is typically 200\,--\,250~mW, and is a mixture
of RF sideband local oscillator power and carrier light resulting from
spatially imperfect interference at the beamsplitter. The light is
divided equally between four length photodetectors, keeping the power
on each at a detectable level of 50\,--\,60~mW. The four length
detector signals are summed and filtered, and the feedback control
signal is applied differentially to the end test masses. This
differential-arm servo loop has a unity-gain bandwidth of
approximately 200~Hz, suppressing fluctuations in the arm lengths to a
residual level of $\sim\!10^{-14}$~m rms. An additional servo is
implemented on these AS port detectors to cancel signals in the
RF-phase orthogonal to the differential-arm channel; this servo
injects RF current at each photodetector to suppress signals that
would otherwise saturate the detectors. About 1\% of the beam is
directed to an alignment detector that controls the differential
alignment of the ETMs.

Maximal power buildup in the interferometer also depends on
maintaining stringent alignment levels. Sixteen alignment
degrees-of-freedom\,--\,pitch and yaw for each of the 6 interferometer
mirrors and the input beam direction\,--\,are controlled by a
hierarchy of feedback loops. First, orientation motion at the pendulum
and isolation stack eigenfrequencies is suppressed locally at each
optic using optical lever angle sensors. Second, global alignment is
established with four RF quadrant photodetectors at the three output
ports as shown in Fig.~\ref{fig:schematic}. These RF alignment
detectors measure wavefront misalignments between the carrier and
sideband fields in a spatial version of PDH detection
\cite{1994ApOpt..33.5041M, 1997JOSAB..14.1597H}. Together the
four detectors provide 5 linearly independent combinations of the
angular deviations from optimal global alignment
\cite{Fritschel:98}. These error signals feed a multiple-input
multiple-output control scheme to maintain the alignment within
$\sim\!10^{-8}$~radians rms of the optimal point, using bandwidths
between $\sim\!0.5$~Hz and $\sim\!5$~Hz depending on the
channel. Finally, slower servos hold the beam centered on the
optics. The beam positions are sensed at the arm ends using DC
quadrant detectors that receive the weak beam transmitted through the
ETMs, and at the corner by imaging the beam spot scattered from the
beamsplitter face with a CCD camera.

The length and alignment feedback controls are all implemented
digitally, with a real-time sampling rate of 16384~samples/sec for the
length controls and 2048~samples/sec for the alignment controls. 
The digital control system provides the flexibility required to
implement the multiple input, multiple output feedback controls
described above.  The digital controls also allow complex filter
shapes to be easily realized, lend the ability to make dynamic
changes in filtering, and make it simple to blend sensor and control
signals.  As an example, optical gain changes are
compensated to first order to keep the loop gains constant
in time by making real-time feed-forward corrections to the digital
gain based on cavity power levels.

The digital controls are also essential to implementing the
interferometer \textit{lock acquisition} algorithm. So far this
section has described how the interferometer is maintained at the
operating point. The other function of the control system is to
acquire lock: to initially stabilize the relative optical positions to
establish the resonance conditions and bring them within the linear
regions of the error signals.  Before lock the suspended optics are
only damped within their suspension structures; ground motion and the
equivalent effect of input-light frequency fluctuations cause the four
(real or apparent) lengths to fluctuate by 0.1\,--\,1~$\mu$m~rms over time
scales of 0.5\,--\,10~sec. The probability of all four
degrees-of-freedom simultaneously falling within the $\sim\!1$~nm
linear region of the resonance points is thus extremely small and a
controlled approach is required. The basic approach of the lock
acquisition scheme, described in detail in reference \cite{Evans:02},
is to control the degrees-of-freedom in sequence: first the
power-recycled Michelson is controlled, then a resonance of one arm
cavity is captured, and finally a resonance of the other arm cavity is
captured to achieve full power buildup. A key element of this scheme
is the real-time, dynamic calculation of a sensor transformation
matrix to form appropriate length error signals throughout the
sequence. The interferometers are kept in lock typically for
many hours at a time, until lock is lost due to environmental
disturbances, instrument malfunction or operator command.

\subsection{Thermal Effects}
At full power operation, a total of 20\,--\,60~mW of light is absorbed
in the substrate and in the mirror surface of each ITM, depending on
their specific absorption levels.  Through the thermo-optic
coefficient of fused silica, this creates a weak, though not
insignificant thermal lens in the ITM substrates
\cite{PhysRevA.44.7022}. Thermo-elastic distortion of the test mass
reflecting surface is not significant at these absorption
levels. While the ITM thermal lens has little effect on the carrier
mode, which is determined by the arm cavity radii of curvature, it
does affect the RF sideband mode supported by the recycling
cavity. This in turn affects the power buildup and mode shape of the
RF sidebands in the recycling cavity, and consequently the sensitivity
of the heterodyne detection signals \cite{D'Ambrosio:2006hq,
  Gretarsson:07}. Achieving maximum interferometer sensitivity thus
depends critically on optimizing the thermal lens and thereby the mode
shape, a condition which occurs at a specific level of absorption in
each ITM (approximately 50~mW). To achieve this optimum mode over the
range of ITM absorption and stored power levels, each ITM thermal lens
is actively controlled by directing additional heating beams,
generated from ${\rm CO_2}$ lasers, onto each ITM
\cite{LIGO:T050064}. The power and shape of the heating beams are
controlled to maximize the interferometer optical gain and
sensitivity. The shape can be selected to have either a Gaussian
radial profile to provide more lensing, or an annular radial profile
to compensate for excess lensing.

\subsection{Interferometer Response and Calibration}

The GW channel is the digital error point of the differential-arm
servo loop. In principle the GW channel could be derived from any
point within this loop. The error point is chosen because the dynamic
range of this signal is relatively small, since the large low-frequency
fluctuations are suppressed by the feedback loop. To calibrate the
error point in strain, the effect of the feedback loop is divided out,
and the interferometer response to a differential arm strain is
factored in \cite{2005CQGra..22S.985L}; this process can be done
either in the frequency domain or directly in the time domain. The
absolute length scale is established using the laser wavelength, by
measuring the mirror drive signal required to move through an
interference fringe. The calibration is tracked during operation with
sine waves injected into the differential-arm loop. The
uncertainty in the amplitude calibration is approximately
$\pm5\%$. Timing of the GW channel is derived from the Global
Positioning System; the absolute timing accuracy of each
interferometer is better than $\pm10\,\mu$sec.

The response of the interferometer output as a function of GW
frequency is calculated in detail in references \cite{Meers:1989rv,
 Fabbro:95, 0264-9381-25-18-184017}. In the long-wavelength
approximation, where the wavelength of the GW is much longer than the
size of the detector, the response $R$ of a Michelson-Fabry-Perot
interferometer is approximated by a single-pole transfer function:
\begin{equation}
   R(f) \propto \frac{1}{1 + if/f_p}\,,
\end{equation}
where the pole frequency is related to the storage time by
$f_p = 1/4 \pi \tau_s$. Above the pole frequency ($f_p = 85$~Hz for 
the LIGO 4~km interferometers), the amplitude response drops off as
$1/f$. As discussed below, the measurement noise above the pole
frequency has a white (flat) spectrum, and so the strain sensitivity
decreases proportionally to frequency in this region. The single-pole
approximation is quite accurate, differing from the exact response
by less than a percent up to $\sim\!1$~kHz \cite{0264-9381-25-18-184017}.

\begin{figure}[t]
\begin{center}
\includegraphics[width=14cm]{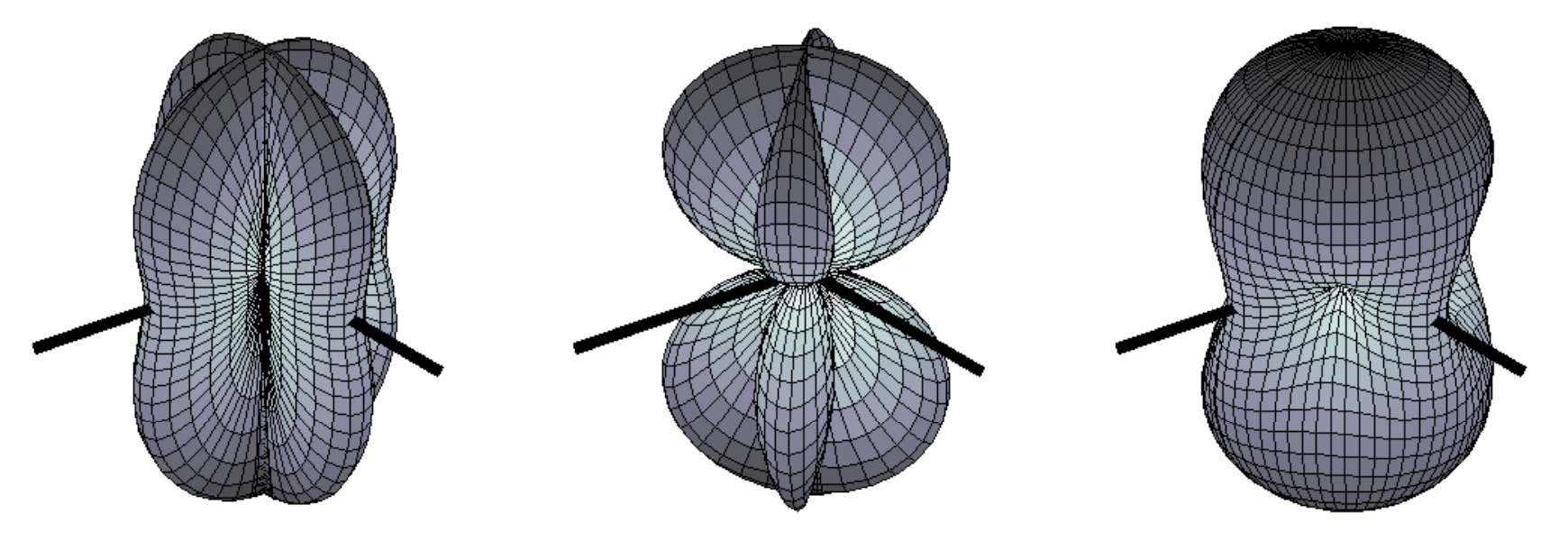}
\caption{Antenna response pattern for a LIGO gravitational wave detector,
in the long-wavelength approximation. 
The interferometer beamsplitter is located at the center of each pattern,
and the thick black lines indicate the orientation of the interferometer 
arms. The distance from a point of the plot surface to the center of the 
pattern is a measure of the gravitational wave sensitivity in this direction. 
The pattern on the left is for $+$ polarization, the middle pattern is for 
$\times$ polarization, and the right-most one is for unpolarized waves. 
\label{fig:antenna}}
\end{center}
\end{figure}

In the long-wavelength approximation, the interferometer directional
response is maximal for GWs propagating orthogonally to the plane of
the interferometer arms, and linearly polarized along the arms. Other
angles of incidence or polarizations give a reduced response, as
depicted by the antenna patterns shown in Fig.~\ref{fig:antenna}. A
single detector has blind spots on the sky for linearly polarized
gravitational waves.

\subsection{Environmental Monitors}
To complete a LIGO detector, the interferometers described above are
supplemented with a set of sensors to monitor the local environment. 
Seismometers and accelerometers measure vibrations of the ground and
various interferometer components; microphones monitor acoustic 
noise at critical locations; magnetometers monitor fields that could
couple to the test masses or electronics; radio receivers monitor
RF power around the modulation frequencies. These sensors are used
to detect environmental disturbances that can couple to the GW channel.

\section{Instrument performance}
\subsection{Strain Noise Spectra} 
\begin{figure}
\begin{center}
\includegraphics[width=12cm]{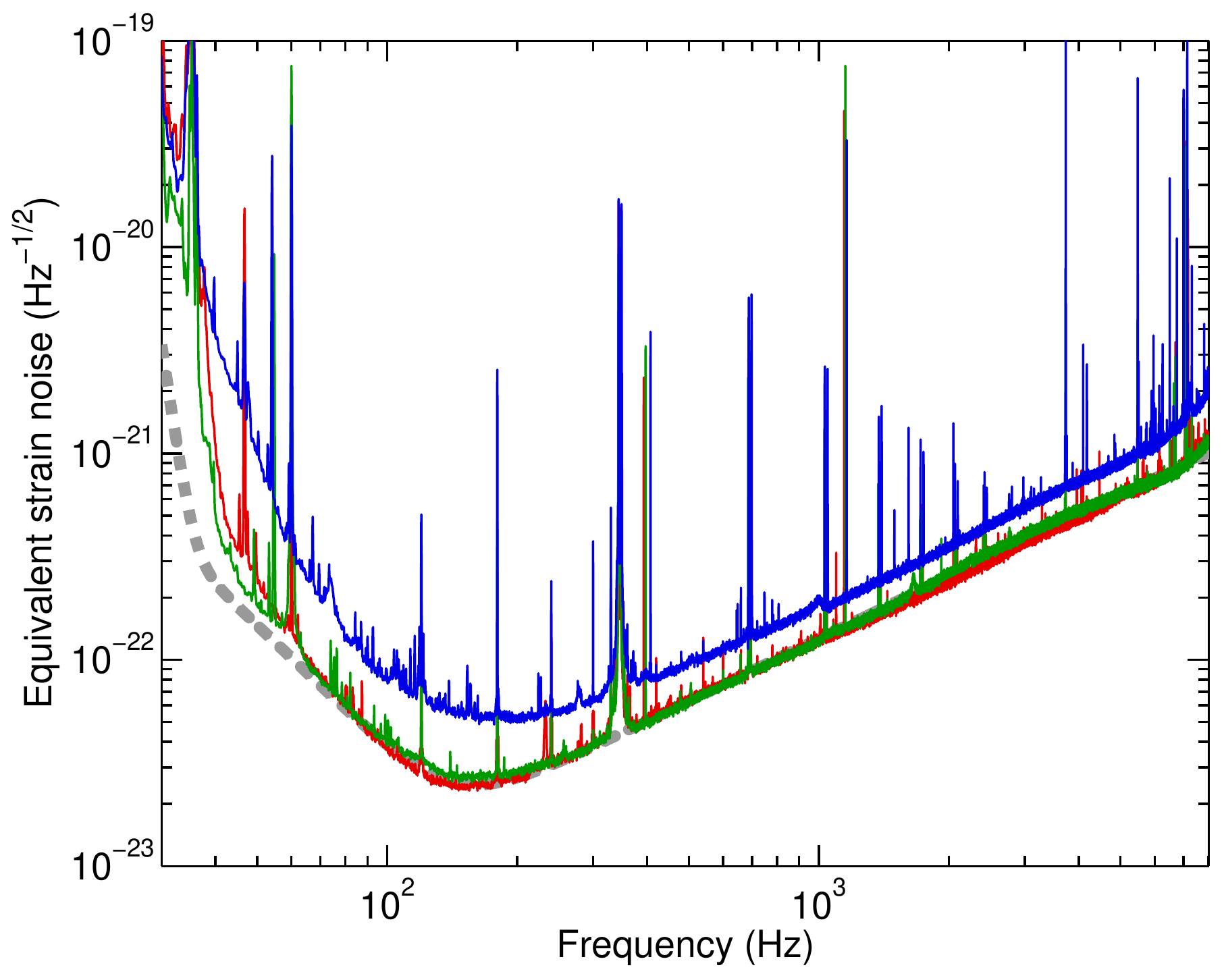}
\caption{Strain sensitivities, expressed as amplitude spectral
  densities of detector noise converted to equivalent GW strain. The
  vertical axis denotes the rms strain noise in 1~Hz of bandwidth.
  Shown are typical high sensitivity spectra for each of the three
  interferometers (red: H1; blue: H2; green: L1), along with the
  design goal for the 4-km detectors (dashed
  grey).\label{fig:strainspectra}}
\end{center}
\end{figure}

During the commissioning period, as the interferometer sensitivity was
improved, several short science runs were carried out, culminating
with the fifth science run (S5) at design sensitivity. The S5 run
collected a full year of triple-detector coincident interferometer
data during the period from November~2005 through
September~2007. Since the interferometers detect GW strain amplitude,
their performance is typically characterized by an amplitude spectral
density of detector noise (the square root of the power spectrum),
expressed in equivalent GW strain. Typical high-sensitivity strain
noise spectra are shown in Fig.~\ref{fig:strainspectra}. Over the
course of S5 the strain sensitivity of each interferometer was
improved, by up to 40\% compared to the beginning of the run through a
series of incremental improvements to the instruments.

\begin{figure}
\begin{center}
\includegraphics[width=11cm]{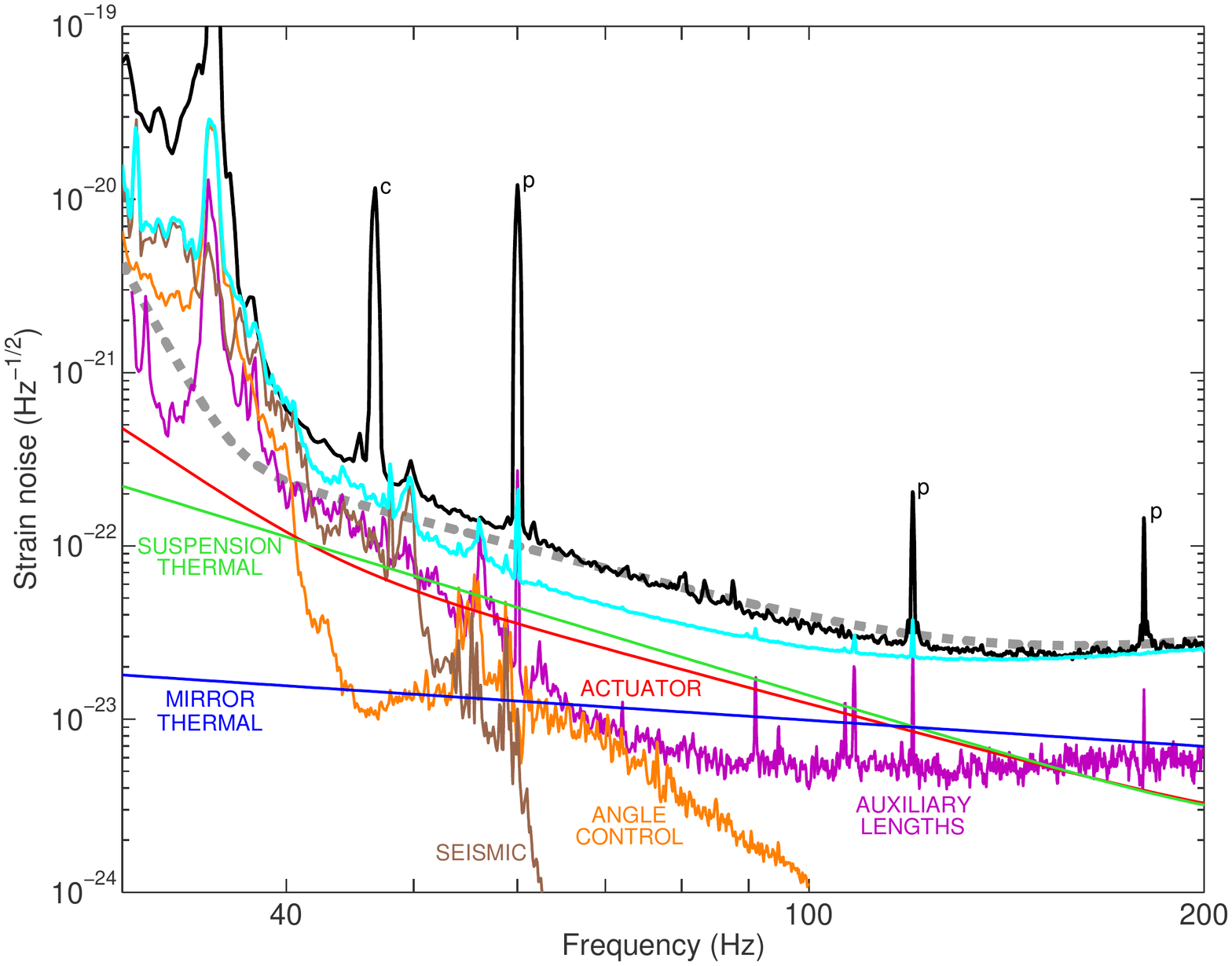}
\includegraphics[width=11cm]{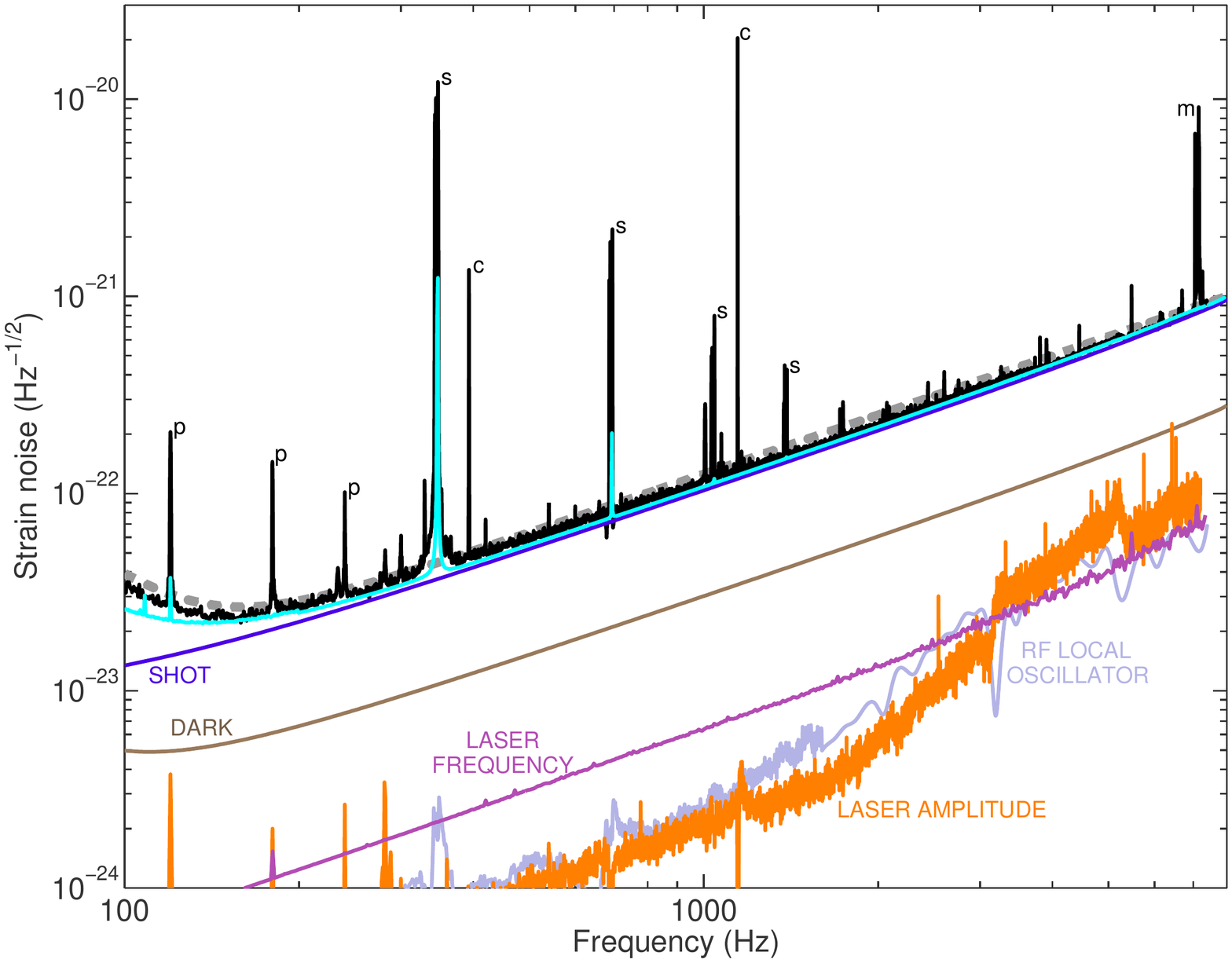}
\caption{Primary known contributors to the H1 detector noise
  spectrum. The upper panel shows the displacement noise components,
  while the lower panel shows sensing noises (note the different
  frequency scales). In both panels, the black curve is the measured
  strain noise (same spectrum as in Fig.~\ref{fig:strainspectra}), the
  dashed gray curve is the design goal, and the cyan curve is the
  root-square-sum of all known contributors (both sensing and
  displacement noises). The labelled component curves are described in
  the text. The known noise sources explain the observed noise very
  well at frequencies above 150~Hz, and to within a factor of 2 in the
  40\,--\,100~Hz band. Spectral peaks are identified as follows: c,
  calibration line; p, power line harmonic; s, suspension wire
  vibrational mode; m, mirror (test mass) vibrational
  mode. \label{fig:noisebudget}}
\end{center}
\end{figure}

The primary noise sources contributing to the H1 strain noise spectrum
are shown in Fig.~\ref{fig:noisebudget}. Understanding and controlling
these instrumental noise components has been the major technical
challenge in the development of the detectors. The noise terms can be
broadly divided into two classes: displacement noise and sensing
noise.  Displacement noises cause motions of the test masses or their
mirrored surfaces. Sensing noises, on the other hand, are phenomena
that limit the ability to measure those motions; they are present even
in the absence of test mass motion.  The strain noises shown in
Fig.~\ref{fig:strainspectra} consists of spectral lines superimposed on
a continuous broadband noise spectrum.  The majority of the lines are
due to power lines ($60, 120, 180, ...$Hz), ``violin mode'' mechanical
resonances ($340, 680, ...$Hz) and calibration lines ($55, 400,$ and $
1100$~Hz).  These high Q lines are easily excluded from analysis; the
broadband noise dominates the instrument sensitivity.

\subsection{Sensing Noise Sources}
Sensing noises are shown in the lower panel of
Fig.~\ref{fig:noisebudget}. By design, the dominant noise source
above $100$~Hz is shot noise, as determined by the Poisson statistics 
of photon detection. The ideal shot-noise limited strain noise density, 
$\tilde{h}(f)$, for this type of interferometer is \cite{Meers:1988wp}:
\begin{equation}\label{snlimit}
    \widetilde{h}(f) = \sqrt{\frac{\pi \hbar \lambda}{\eta P_{\rm BS} c}}\frac{\sqrt{1+(4 \pi f \tau_{\rm s})^2}}{4 \pi
    \tau_{\rm s}},
\end{equation}
where $\lambda$ is the laser wavelength, $\hbar$ is the reduced Planck
constant, $c$ is the speed of light, $\tau_{\rm s}$ is the arm cavity
storage time, $f$ is the GW frequency, $P_{\rm BS}$ is the power
incident on the beamsplitter, and $\eta$ is the photodetector quantum
efficiency. For the estimated effective power of $\eta P_{\rm BS} =
0.9\cdot250$~W, the ideal shot-noise limit is $\widetilde{h} = 1.0
\times 10^{-23}/\sqrt{{\rm Hz}}$ at 100~Hz. The shot-noise estimate in
Fig.~\ref{fig:noisebudget} is based on measured photocurrents in the
AS port detectors and the measured interferometer response. The
resulting estimate, $\widetilde{h}(100 {\rm Hz}) = 1.3 \times
10^{-23}/\sqrt{{\rm Hz}}$, is higher than the ideal limit due to
several inefficiencies in the heterodyne detection process: imperfect
interference at the beamsplitter increases the shot noise; imperfect
modal overlap between the carrier and RF sideband fields decreases the
signal; and the fact that the AS port power is modulated at twice the
RF phase modulation frequency leads to an increase in the
time-averaged shot noise \cite{PhysRevA.43.5022}.

Many noise contributions are estimated using stimulus-response tests,
where a sine-wave or broadband noise is injected into an auxiliary
channel to measure its coupling to the GW channel. This method is used
for the laser frequency and amplitude noise estimates, the RF
oscillator phase noise contribution, and also for the angular control
and auxiliary length noise terms described below. Although laser noise
is nominally common-mode, it couples to the GW channel through small,
unavoidable differences in the arm cavity mirrors \cite{LIGO:T970084,
  2000JOSAA..17..120C}. Frequency noise is expected to couple most
strongly through a difference in the resonant reflectivity of the two
arms. This causes carrier light to leak out the AS port, which
interferes with frequency noise on the RF sidebands to create a noise
signal. The stimulus-response measurements indicate the coupling is
due to a resonant reflectivity difference of about 0.5\%, arising from
a loss difference of tens of ppm between the arms. Laser amplitude
noise can couple through an offset from the carrier dark fringe. The
measured coupling is linear, indicating an effective static offset of
$\sim\!1$~picometer, believed to be due to mode shape differences
between the arms.

\subsection{Seismic and Thermal Noise}
Displacement noises are shown in the upper panel of
Fig.~\ref{fig:noisebudget}. At the lowest frequencies the largest
such noise is seismic noise\,--\,motions of the earth's surface
driven by wind, ocean waves, human activity, and low-level
earthquakes\,--\,filtered by the isolation stacks and pendulums. The
seismic contribution is estimated using accelerometers to measure
the vibration at the isolation stack support points, and propagating
this motion to the test masses using modeled transfer functions of
the stack and pendulum. The seismic wall frequency, below which
seismic noise dominates, is approximately 45~Hz, a bit higher than
the goal of 40~Hz, as the actual environmental vibrations around
these frequencies are $\sim\!10$ times higher than was estimated in
the design.

Mechanical thermal noise is a more fundamental effect, arising from
finite losses present in all mechanical systems, and is governed by
the fluctuation-dissipation theorem \cite{Saulson:1990jc,
  Levin:1997kv}. It causes arm length noise through thermal excitation
of the test mass pendulums (\textit{suspension thermal noise})
\cite{0264-9381-17-21-305}, and thermal acoustic waves that perturb
the test mass mirror surface (\textit{test mass thermal noise})
\cite{0264-9381-19-5-305}. Most of the thermal energy is concentrated
at the resonant frequencies, which are designed (as much as possible)
to be outside the detection band. Away from the resonances, the level
of thermal motion is proportional to the mechanical dissipation
associated with the motion. Designing the mirror and its pendulum to
have very low mechanical dissipation reduces the detection-band
thermal noise. It is difficult, however, to accurately and
unambiguously establish the level of broadband thermal noise
\textit{in-situ}; instead, the thermal noise curves in
Fig.~\ref{fig:noisebudget} are calculated from models of the
suspension and test masses, with mechanical loss parameters taken from
independent characterizations of the materials.

For the pendulum mode, the mechanical dissipation occurs near the
ends of the suspension wire, where the wire flexes. Since the
elastic energy in the flexing regions depends on the wire radius to
the fourth power, it helps to make the wire as thin as possible to limit
thermal noise. The pendulums are thus made with steel wire for its
strength; with a diameter of 300~$\mu$m the wires are loaded to 30\% of
their breaking stress. The thermal noise in the pendulum mode of the
test masses is estimated assuming a frequency-independent
mechanical loss angle in the suspension wire of $3 \times 10^{-4}$
\cite{Gillespie:1994fg}. This is a relatively small loss for a metal wire
\cite{Cagnoli1999230}.

Thermal noise of the test mass surface is associated with mechanical
damping within the test mass. The fused-silica test mass substrate
material has very low mechanical loss, of order $10^{-7}$ or smaller
\cite{Penn20063}. On the other hand, the thin-film, dielectric
coatings that provide the required optical
reflectivity\,--\,alternating layers of silicon dioxide and tantalum
pentoxide\,--\, have relatively high mechanical loss. Even though the
coatings are only a few microns thick, they are the dominant source of
the relevant mechanical loss, due to their level of dissipation and
the fact that it is concentrated on the test mass face probed by the
laser beam \cite{Levin:1997kv}. The test mass thermal noise estimate
is calculated by modeling the coatings as having a
frequency-independent mechanical dissipation of $4 \times 10^{-4}$
\cite{0264-9381-19-5-305}.

\subsection{Auxiliary Degree-of-freedom Noise}
The auxiliary length noise term refers to noise in the Michelson and
power recycling cavity servo loops which couple to the GW channel. The
former couples directly to the GW channel while the latter couples in
a manner similar to frequency noise.  Above $\sim\!50$~Hz the
sensing noise in these loops is dominated by shot noise; since loop
bandwidths of $\sim\!100$~Hz are needed to adequately stabilize these
degrees of freedom, shot noise is effectively added onto their
motion. Their noise infiltration to the GW channel, however, is
mitigated by appropriately filtering and scaling their digital control
signals and adding them to the differential-arm control signal as a
type of feed-forward noise suppression \cite{Fritschel:01}. These
correction paths reduce the coupling to the GW channel by
10\,--\,40~dB.

We illustrate this more concretely with the Michelson loop. The
shot-noise-limited sensitivity for the Michelson is $\sim\!10^{-16}\,
{\rm m/\sqrt{Hz}}$.  Around 100~Hz, the Michelson servo impresses this
sensing noise onto the Michelson degree-of-freedom (specifically, onto
the beamsplitter).  Displacement noise in the Michelson couples to
displacement noise in the GW channel by a factor of $\pi /(\sqrt{2} F) =
1/100$, where $F$ is the arm cavity finesse. The Michelson sensing
noise would thus produce $\sim\!10^{-18}\,{\rm m/\sqrt{Hz}}$ of GW
channel noise around 100~Hz, if uncorrected. The digital correction
path subtracts the Michelson noise from the GW channel with
an efficiency of 95\% or more.  This brings the Michelson noise
component down to $\sim\!10^{-20}\,{\rm m/\sqrt{Hz}}$ in the GW
channel, 5\,--\,10 times below the GW channel noise floor.

Angular control noise arises from noise in the alignment sensors (both
optical levers and wavefront sensors), propagating to the test masses
through the alignment control servos. Though these feedback signals
affect primarily the test mass orientation, there is always some
coupling to the GW degree-of-freedom because the laser beam is not
perfectly aligned to the center-of-rotation of the test mass surface
\cite{Kawamura:94}. Angular control noise is minimized by a
combination of filtering and parameter tuning. Angle control
bandwidths are 10~Hz or less and strong low-pass filtering is applied
in the GW band. In addition, the angular coupling to the GW channel is
minimized by tuning the center-of-rotation, using the four actuators
on each optic, down to typical residual coupling levels of
$10^{-3}-10^{-4}$~m/rad.

\subsection{Actuation Noise}
The actuator noise term includes the electronics that produce the coil
currents keeping the interferometer locked and aligned, starting with
the digital-to-analog converters (DACs). The actuation electronics
chain has extremely demanding dynamic range requirements. At low
frequencies, control currents of $\sim\!10$~mA are required to provide
$\sim\!5\,\mu$m of position control, and tens of mA are
required to provide $\sim\!0.5$~mrad of alignment bias. Yet the
current noise through the coils must be kept below a couple of
pA$/\sqrt{\rm Hz}$ above 40~Hz. The relatively limited dynamic
range of the DACs is managed with a combination of digital and analog
filtering: the higher frequency components of the control signals are
digitally emphasized before being sent to the DACs, and then
de-emphasized following the DACs with complementary analog filters.
The dominant coil current noise comes instead from the circuits that
provide the alignment bias currents, followed closely by the circuits
that provide the length feedback currents.

\subsection{Additional Noise Sources}
In the 50\,--\,100~Hz band, the known noise sources typically do not
fully explain the measured noise. Additional noise mechanisms have
been identified, though not quantitatively established. Two
potentially significant candidates are nonlinear conversion of low
frequency actuator coil currents to broadband noise (upconversion),
and electric charge build-up on the test masses. A variety of
experiments have shown that the upconversion occurs in the magnets
(neodymium iron boron) of the coil-magnet actuators, and produces a
broadband force noise, with a $f^{-2}$ spectral slope; this is the
phenomenon known as Barkhausen noise \cite{Cote:1991p8320}. The
nonlinearity is small but not negligible given the dynamic range
involved: 0.1~mN of low-frequency (below a few Hertz) actuator force
upconverts of order $10^{-11}$~N rms of force noise in the
40\,--\,80~Hz octave. This noise mechanism is significant primarily
below 80~Hz, and varies in amplitude with the level of ground motion
at the observatories.   

Regarding electric charge, mechanical contact of a test mass with its
nearby limit-stops, as happens during a large earthquake, can build up
charge between the two objects. Such charge distributions are not
stationary; they tend to redistribute on the surface to reduce local
charge density. This produces a fluctuating force on the test mass,
with an expected $f^{-1}$ spectral slope. Although the level at which
this mechanism occurs in the interferometers is not well-known,
evidence for its potential significance comes from a fortuitous event
with L1. Following a vacuum vent and pump-out cycle partway through
the S5 science run, the strain noise in the 50\,--\,100~Hz band went
down by about 20\%; this was attributed to charge reduction on one of
the test masses.

In addition to these broadband noises, there are a variety of periodic
or quasi-periodic processes that produce lines or narrow features in
the spectrum. The largest of these spectral peaks are identified in
Fig.~\ref{fig:noisebudget}. The groups of lines around 350~Hz, 700~Hz,
\textit{et cetera} are vibrational modes of the wires that suspend the
test masses, thermally excited with $kT$ of energy in each mode. The
power line harmonics, at 60~Hz, 120~Hz, 180~Hz, \textit{et cetera}
infiltrate the interferometer in a variety of ways. The 60~Hz line,
for example, is primarily due to the power line's magnetic field
coupling directly to the test mass magnets. As all these lines are
narrow and fairly stable in frequency, they occupy only a small
fraction of the instrument spectral bandwidth.

\subsection{Other Performance Figures-of-merit}
\begin{figure}
\begin{center}
\includegraphics[width=12cm]{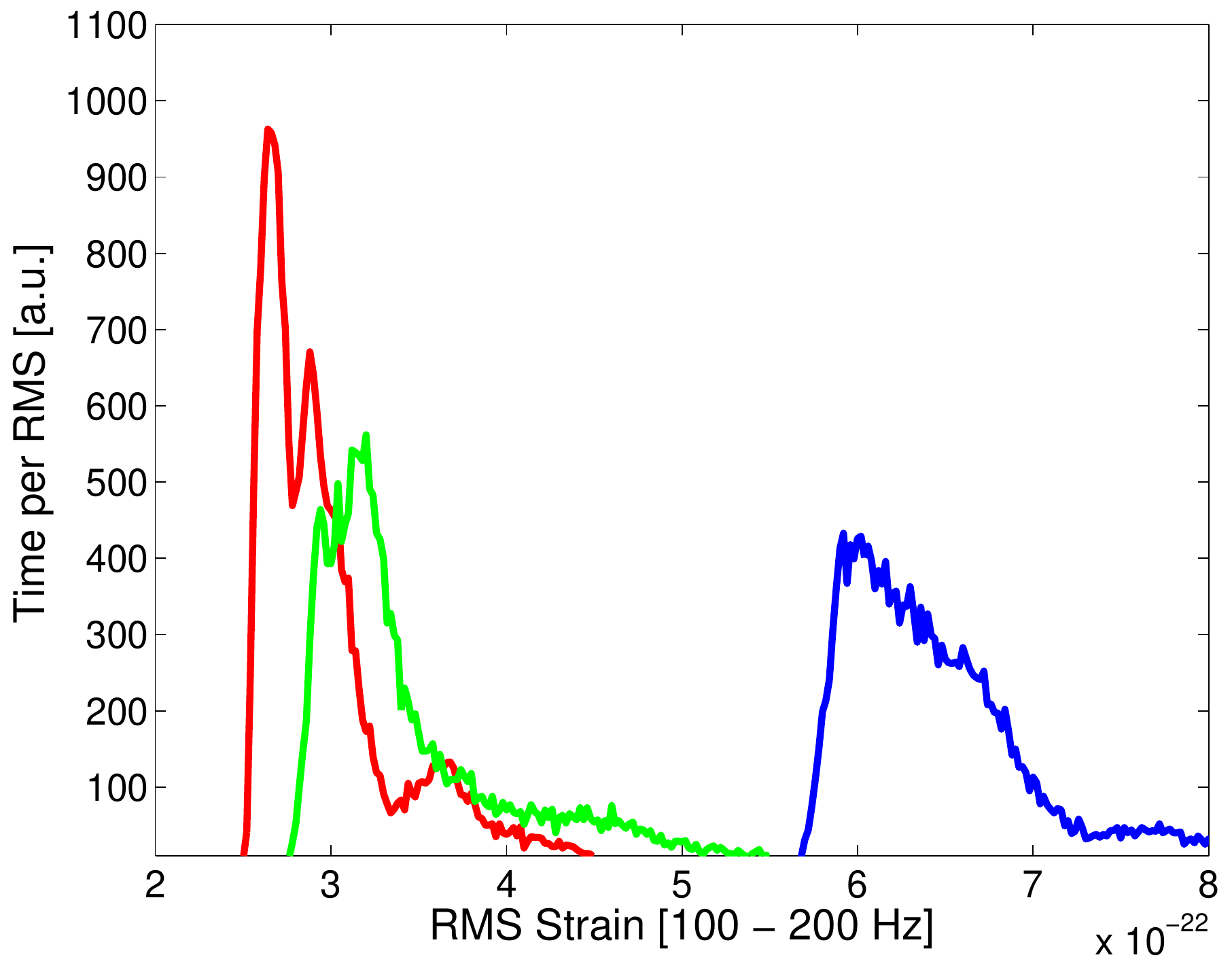}
\caption{Histograms of the RMS strain noise in the band $100-200$~Hz,
computed from the S5 data for each of the LIGO interferometers 
(red: H1; green: L1; blue: H2). Each RMS strain value is calculated
using 30~minutes of data. Much of the higher RMS portions of each
distribution date from the first $\sim\!100$ days of the run,
around which time sensitivity improvements were made to all 
interferometers. Typical RMS variations over daily and weekly
time scales are $\pm$5\% about the mean. With the half arm-length
of H2, its RMS strain noise in this band is expected to be about
two times higher than that of H1 and L1.\label{fig:RMShist}}
\end{center}
\end{figure}

While Figs.~\ref{fig:strainspectra} and \ref{fig:noisebudget} show
high-sensitivity strain noise spectra, the interferometers exhibit
both long- and short-term variation in sensitivity due to improvements
made to the detectors, seasonal and daily variations in the
environment, and the like. One indicator of the sensitivity variation
over the S5 science run is shown in Fig.~\ref{fig:RMShist}: histograms
of the rms strain noise in the frequency band of 100\,--\,200~Hz.

\begin{figure}
\begin{center}
\includegraphics[width=12cm]{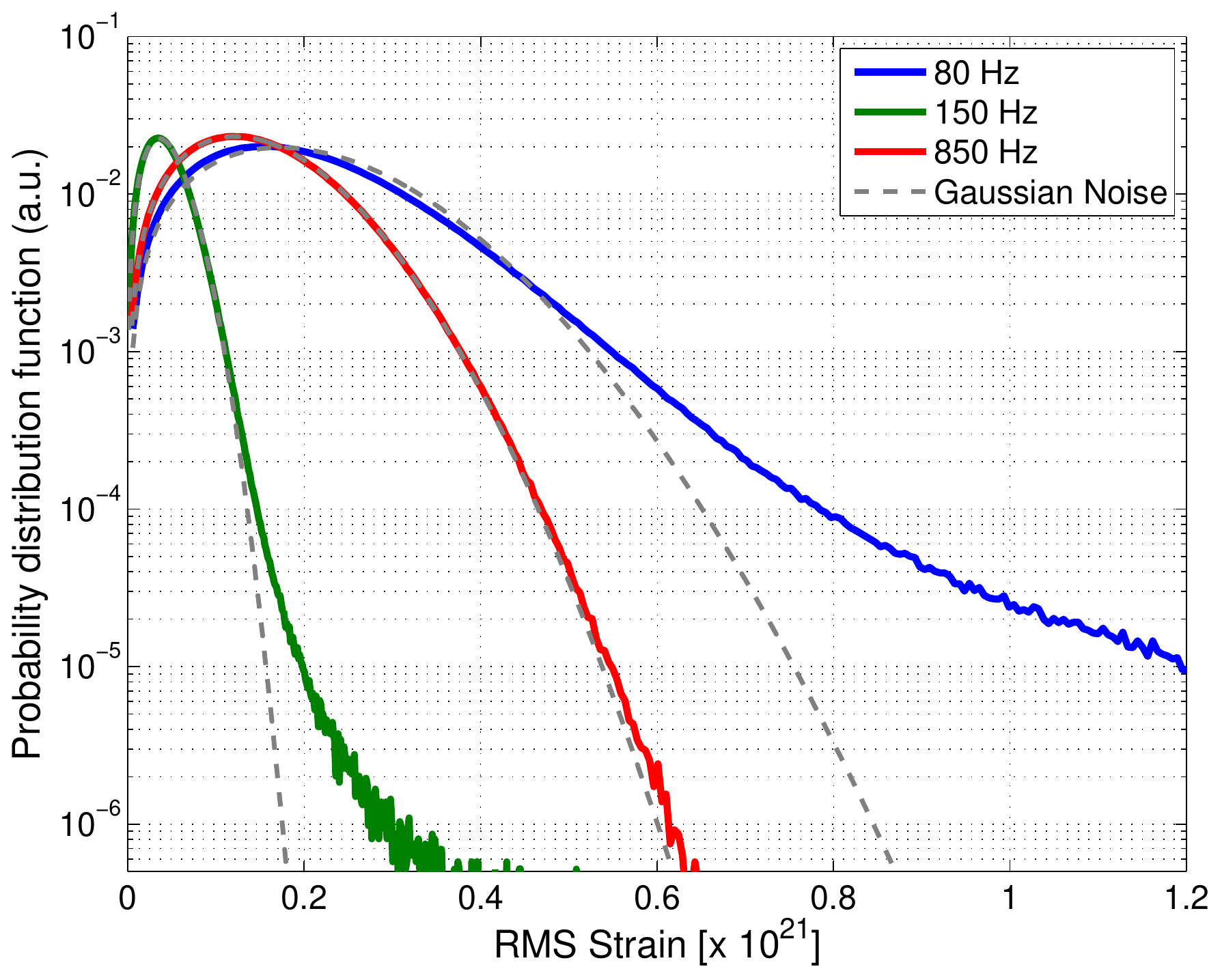}
\caption{Distribution of strain noise amplitude for three
  representative frequencies within the measurement band (data shown
  for the H1 detector). Each curve is a histogram of the spectral
  amplitude at the specified frequency over the second half of the S5
  data run. Each spectral amplitude value is taken from the Fourier
  transform of 1 second of strain data; the equivalent noise bandwidth
  for each curve is 1.5~Hz. For comparison, the dashed grey lines are
  Rayleigh distributions, which the measured histograms would follow
  if they exhibited stationary, Gaussian noise. The high frequency
  curve is close to a Rayleigh distribution, since the noise there is
  dominated by shot noise. The lower frequency curves, on the other
  hand, exhibit non-Gaussian fluctuations.\label{fig:Rayleighplot}}
\end{center}
\end{figure}

To get a sense of shorter term variations in the noise, 
Fig.~\ref{fig:Rayleighplot} shows the distribution of strain
noise amplitudes at three representative frequencies where the noise
is dominated by random processes. For stationary, Gaussian noise
the amplitudes would follow a Rayleigh distribution, and
deviations from that indicate non-Gaussian fluctuations. 
As Fig.~\ref{fig:Rayleighplot} suggests, the lower frequency
end of the measurement band shows a higher level of non-Gaussian
noise than the higher frequencies. Some of this non-Gaussianity
is due to known couplings to a fluctuating environment; much
of it, however, is due to glitches\,--\,any short duration
noise transient\,--\,from unknown mechanisms. Additional
characterizations of the glitch behavior of the detectors
can be found in reference \cite{0264-9381-25-18-184004}.

Another important statistical figure-of-merit is the interferometer
duty cycle, the fraction of time that detectors are operating and
taking science data. Over the S5 period, the individual interferometer
duty cycles were 78\%, 79\%, and 67\% for H1, H2, and L1,
respectively; for double-coincidence between L1 and H1 or H2
the duty cycle was 60\%; and for triple-coincidence of all three
interferometers the duty cycle was 54\%. These figures include
scheduled maintenance and instrument tuning periods, as well as
unintended losses of operation.

\section{Data Analysis Infrastructure}

While the LIGO interferometers provide extremely sensitive
measurements of the strain at two distant locations, the instruments
constitute only one half of the ``Gravitational-wave Observatory'' in
LIGO.  The other half is the computing infrastructure and data
analysis algorithms required to pull out gravitational wave signals
from the noise. Potential sources and the methods used to search for
them are discussed in the next section. First, we discuss some
features of the LIGO data and their analysis that are common to all
searches.

The raw instrument data are collected and archived for off-line
analysis. For each detector, approximately $50$ channels are recorded
at a sample rate of 16,384~Hz, $550$ channels at reduced rates of 256
to 4,096~Hz, and $6000$ digital monitors at 16~Hz. The aggregate rate
of archived data is about 5~MB/s for each interferometer. Computer
clusters at each site also produce reduced data sets containing only
the most important channels for analysis.

The detector outputs are pre-filtered with a series of data quality
checks to identify appropriate time periods to analyze. The most
significant data quality (DQ) flag, ``science mode'', ensures the
detectors are in their optimum run-time configuration; it is set by
the on-site scientists and operators.  Follow-up DQ flags are set for
impending lock loss, hardware injections, site disturbances, and data
corruptions. DQ flags are also used to mark times when the instrument
is outside its nominal operating range, for instance when a sensor or
actuator is saturating, or environmental conditions are unusually
high. Depending on the specific search algorithm, the DQ flags
introduce an effective dead-time of 1\% to 10\% of the total science
mode data.

Injections of simulated gravitational wave signals are performed to test
the functionality of all the search algorithms and also to measure
detection efficiencies. These injections are done both in software,
where the waveforms are added to the archived data stream, and directly
in hardware, where they are added to the feedback control signal in
the differential-arm servo. In general the injected waveforms simulate
the actual signals being searched for, with representative waveforms
used to test searches for unknown signals.

As described in the section on instrument performance, the local
environment and the myriad interferometer degrees-of-freedom can all
couple to the gravitational wave channel, potentially creating
artifacts that must be distinguished from actual
signals. Instrument-based vetoes are developed and used to reject such
artifacts \cite{0264-9381-25-18-184004}. The vetoes are tested using
hardware injections to ensure their safety for gravitational wave
detections. The efficacy of these vetoes depends on the search type.

\section{Astrophysical Reach and Search Results}
LIGO was designed so that its data could be searched for GWs from many
different sources. The sources can be broadly characterized as either
transient or continuous in nature, and for each type, the analysis
techniques depend on whether the gravitational waveforms can be
accurately modeled or whether only less specific spectral
characterizations are possible. We therefore organize the 
searches into four categories according to source type
and analysis technique:

\begin{enumerate}
\item{Transient, modeled waveforms:} the
  \textit{compact binary coalescence} search. The name follows from
  the fact that the best understood transient sources are the final
  stages of binary inspirals \cite{Belczynski:2001uc}, where each
  component of the binary may be a neutron star (NS) or a black hole
  (BH). For these sources the waveform can be calculated with good
  precision, and matched-filter analysis can be used.

\item{Transient, unmodeled waveforms:} the
  \textit{gravitational-wave bursts} search. Transient systems such as
  core-collapse supernovae \cite{Dimmelmeier:2001rw}, black-hole
  mergers, and neutron star quakes, may produce GW bursts that can
  only be modeled imperfectly, if at all, and more general analysis
  techniques are needed.

\item{Continuous, narrow-band waveforms:} the
  \textit{continuous wave sources} search. An example of a continuous
  source of GWs with a well-modeled waveform is a spinning neutron
  star (e.g., a pulsar) that is not perfectly symmetric about its
  rotation axis \cite{Jaranowski:1998qm}.

\item{Continuous, broad-band waveforms:} the
  \textit{stochastic gravitational-wave background} search. Processes
  operating in the early universe, for example, could have produced a
  background of GWs that is continuous but stochastic in character
  \cite{Maggiore:1999vm}.
\end{enumerate}

In the following sections we review the astrophysical results that
have been generated in each of these search categories using LIGO
data; reference \cite{LIGOpubs} contains links to all the LIGO
observational publications. To date, no GW signal detections have been
made, so these results are all upper limits on various GW sources. In
those cases where the S5 analysis is not yet complete, we present the
most recent published results and also discuss the expected
sensitivity, or astrophysical reach, of the search based on the S5
detector performance.

\subsection{Compact Binary Coalescences}
Binary coalescences are unique laboratories for testing general
relativity in the strong-field regime \cite{lrr-2006-3}. GWs from such
systems will provide unambiguous evidence for the existence of black
holes and powerful tests of their properties as predicted by general
relativity \cite{lrr-2006-4, lrr-2003-6}. Multiple observations will
yield valuable information about the population of such systems in the
universe, up to distances of hundreds of megaparsecs (Mpc, 1 parsec =
3.3 light years).  Coalescences involving neutron stars will provide
information about the nuclear equation of state in these extreme
conditions. Such systems are considered likely progenitors of
short-duration gamma ray bursts (GRBs) \cite{2005Natur.437..845F}.

Post-Newtonian approximations to general relatively accurately model a
binary system of compact objects whose orbit is adiabatically
tightening due to GW emission \cite{Blanchet:2002av}. Several examples
of such binary systems exist with merger times less than the age of
the universe, most notably the binary pulsar system PSR 1913+16 described
previously. After an extended \textit{inspiral} phase, the system
becomes dynamically unstable when the separation decreases below an
innermost stable circular orbit (approximately 25~km for two
solar-mass neutron stars) and the objects plunge and form a single
black hole in the \textit{merger} phase.  The resulting distorted
black hole relaxes to a stationary Kerr state via the strongly damped
sinusoidal oscillations of the quasi-normal modes in the
\textit{ringdown} phase.  The smoothly evolving inspiral and ringdown
GW waveforms can be approximated analytically, while the extreme
dynamics of the merger phase require numeric solutions to determine
the GW waveform \cite{Flanagan:1997sx}.  Collectively, the inspiral,
merger and ringdown of a binary system is termed a Compact Binary
Coalescence (CBC).

The waveform for a compact binary inspiral is a chirp: a sinusoid
increasing in frequency and amplitude until the end of the inspiral
phase. The inspiral phase of a neutron star binary (BNS, with each
mass assumed to be 1.4 $M_\odot$) will complete nearly 2,000 orbits in
the LIGO band over tens of seconds before merger, and emit a maximum
GW frequency of about 1500~Hz. Higher mass inspirals terminate at
proportionally lower GW frequencies. For non-spinning objects, the
inspiral waveform is uniquely determined by the two component masses
$m_1$ and $m_2$ of the system \cite{Blanchet:1995ez}. No analytic
waveforms exist for the merger phase; calculating these waveforms is
one of the primary goals of numerical relativity \cite{Hannam:2009rd,
  Aylott:2009ya}. The ringdown phase is described by an
exponentially-damped sinusoid, determined by the quasi-normal mode
frequency and the quality factor of the final black hole
\cite{Creighton:1999pm}.

\subsubsection{Analysis method}

Since the inspiral and ringdown waveforms for a given mass pair $(m_1,
m_2)$ are accurately known, searches for them are performed using
optimal matched filtering employing a bank of templates covering the
desired $(m_1, m_2)$ parameter space.  An optimized algorithm
generates the template bank, minimizing the number of templates while
allowing a maximum Signal to Noise Ratio (SNR) loss of 3\%
\cite{Owen:1998dk,Babak:2006ty,Cokelaer:2007kx}. In practice
approximately seven thousand templates are used to cover total masses
between 2 and 35 $M_\odot$.

The matched filtering process generates a trigger when the SNR of the
filter output exceeds a threshold. The threshold is set by balancing
two factors: it must be low enough so that a good estimation can be
made of the background due to detector noise, and it must be high
enough to keep the number of triggers manageable.  Associated with
each trigger is a specific template, or mass pair, and a coalescence
time which maximize the SNR for that signal event
\cite{Abbott:2007xi}.

Triggers are first generated independently for each detector. The
number of false triggers created by detector noise is then greatly
reduced by finding the set of \textit{coincident triggers}\,--\,those
that correspond to similar template masses and coalescence times,
within appropriate windows, between at least two LIGO
detectors. Coincident triggers are subject to additional consistency
checks, such as the $\chi^2$ \cite{Allen:2004gu} and $r^2$
\cite{Rodriguez:2007} tests.

Typically many thousands of coincident triggers per month remain at
the end of the pipeline. These surviving triggers are compared with
the background from accidental coincidences of triggers due to
detector noise. Time shift trials are used to estimate the background:
the analysis is repeated with the triggers from different detectors
shifted in time relative to each other by an amount large compared to
the coincidence window. A hundred such trials are typically made. For
each region of mass parameter space, the time shift trials establish a
false alarm rate as a function of SNR. In-time coincident triggers
with the smallest false alarm rate are potential detection candidates
\cite{Collaboration:2009tt}.

A large number of software injections is made to tune the analysis
pipeline and evaluate its detection efficiency.  The injected
waveforms cover the largest practical range of parameter space
possible (component masses, spins, orientations, sky locations and
distances).  The resulting detection efficiency is combined with
simple models of the astrophysical source distribution to arrive at an
estimated cumulative luminosity to which the search is sensitive.
These models \cite{Kopparapu:2008, Kalogera:2003tn} predict that the
rate of CBCs should be proportional to the stellar birth rate in
nearby spiral galaxies. This birth rate can be estimated from a
galaxy's blue luminosity\footnote{Blue luminosity is short for 
B-band luminosity, signifying one of a standard set of optical 
filters used in measuring the luminosity of galaxies.}, 
so we express the cumulative luminosity in
units of $L_{10}$, where $L_{10}$ is $10^{10}$ times the blue solar
luminosity (the Milky Way contains $\sim\!1.7 L_{10}$).

\subsubsection{Analysis Results}
To date, the detection candidates resulting from the analysis pipeline
are consistent with the estimated background and thus are likely
accidental coincidences.  In the absence of detection, mass-dependent
upper limits are set on the rate of CBCs in the local universe. These
rate limits are expressed per unit $L_{10}$.

\begin{figure}[hbt]
  \centering
  \includegraphics[width=4in]{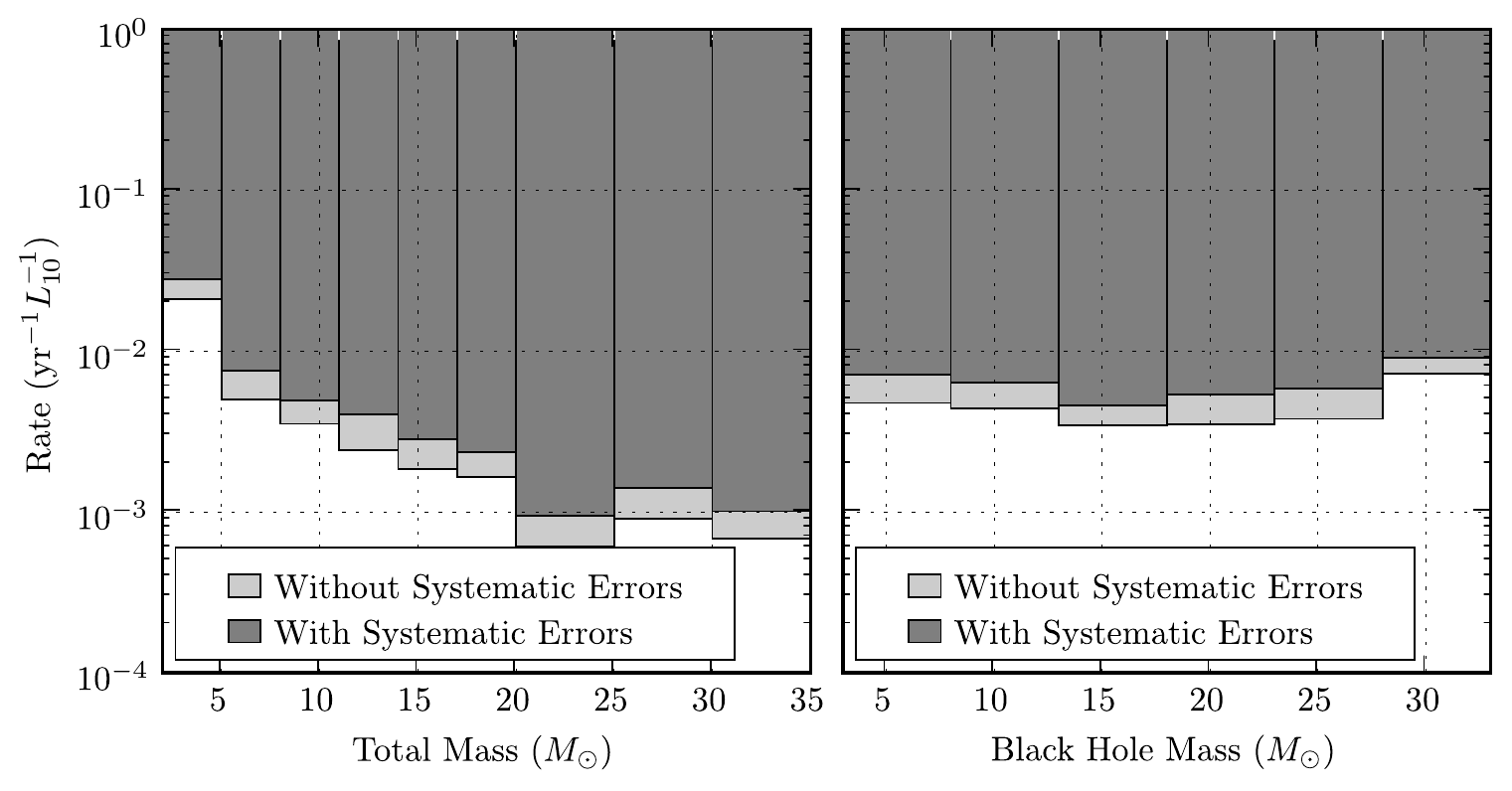}
  \caption{S5 year 1 upper limits on the binary coalescence rate per year 
   and per L10 as a function of total mass of the binary system assuming a uniform 
   distribution in the mass ratio (left) and as a function of the mass of a 
   black hole in a BHNS system with a neutron star mass of $1.35 M_\odot$
   (right). The darker area shows the excluded region when accounting for 
   marginalization over systematic errors. The lighter area shows the additional 
   region that would have been excluded if systematic errors 
   had been ignored. From reference \cite{Collaboration:2009tt}.}
  \label{fig:s5yr1rates}
\end{figure}

An inspiral search with total masses between 2 and 35 $M_\odot$
has been completed using the first calendar year of S5 data
\cite{Collaboration:2009tt}.  Figure \ref{fig:s5yr1rates} shows the
resulting rate upper limit for low mass binary coalescences as a
function of the total mass (left), and as a function of the mass of a
black hole in a black hole-neutron star system with a neutron star
mass of $1.35 M_\odot$ (right). The same analysis set a binary neutron
star coalescence rate upper limit of $3.8 \times 10^{-2}\,yr^{-1}
L_{10}^{-1}$ . This upper limit is still significantly higher than
recent CBC rate estimates derived from the observed BNS
population\,--\,approximately $5 \times 10^{-5}\,yr^{-1} L_{10}^{-1}$
for NS/NS binaries \cite{Kalogera:2003tn}.

Since the LIGO sensitivities improved as S5 progressed, analysis of
the full data set should provide significantly more interesting
coalescence results. In the meantime, the astrophysical reach for
these sources can be estimated from the detector noise performance.
The minute-by-minute strain noise spectra for each detector are used
to calculate the \textit{horizon distance}: the maximum distance at
which an inspiral could be detected with an SNR of 8. For BNS
inspirals, the horizon distance was 30\,--\,35~Mpc each for L1 and H1,
and about 17~Mpc for H2. Based on the increased horizon distances and
extrapolations from the first-year search results, we expect to
achieve better than a factor of two increase in cumulative exposure
with the full run analysis.



\begin{figure}[hbt]
  \centering
  \includegraphics[width=4in]{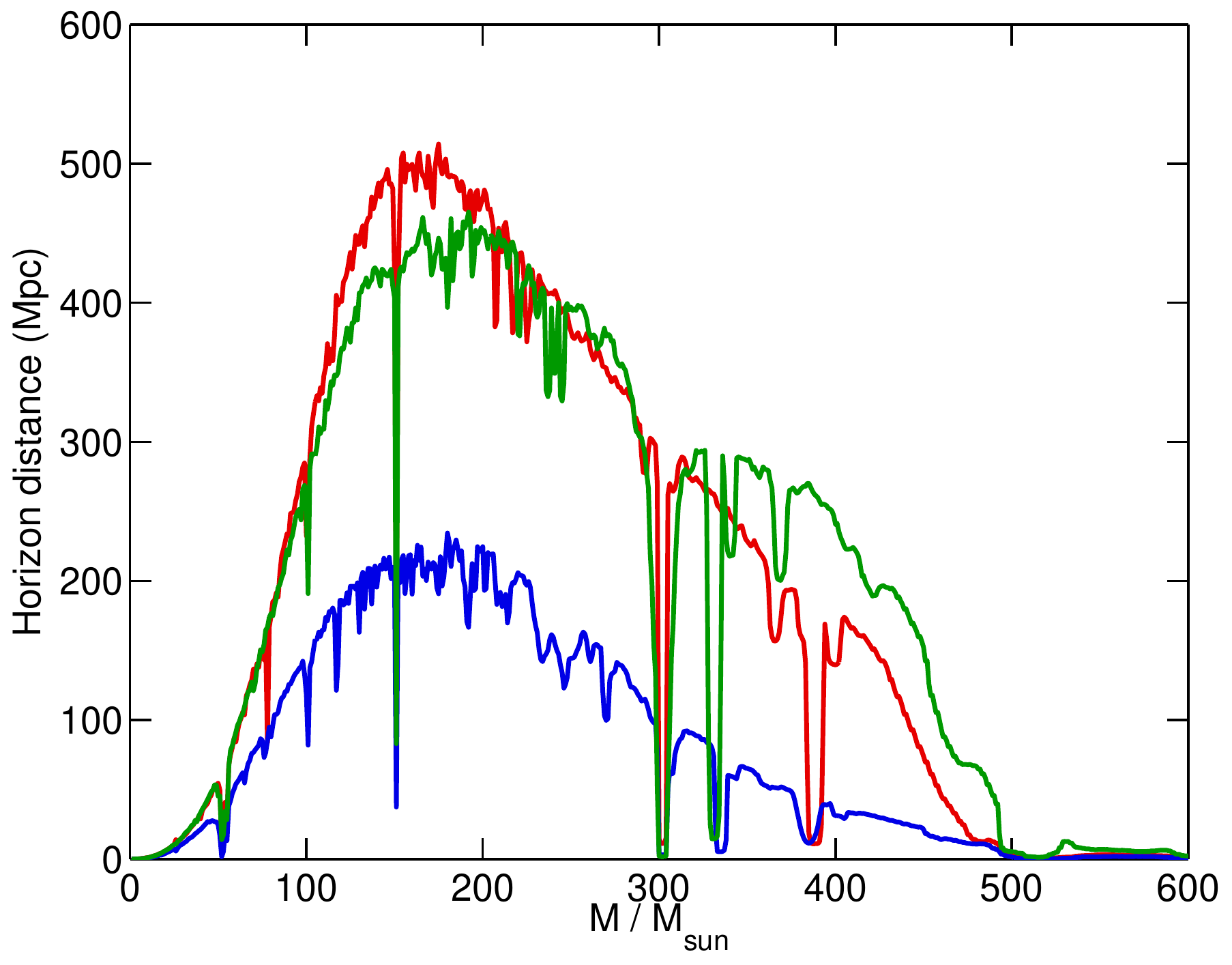}
  \caption{S5 sensitivity to binary black hole ringdowns for the H1 (red),
   L1 (green) and H2 (blue) detectors.  When the ringdown frequency
   coincides with a spectral line the sensitivity is much reduced (300
   $\mathrm{M/M_{sun}}$ corresponds to 60~Hz).}
  \label{fig:blackholes}
\end{figure}

The sensitivity to black hole ringdowns is similarly estimated using
the S5 detector strain noise. Figure~\ref{fig:blackholes} shows the
single detector range for black hole ringdowns averaged over sky
position and spin orientation. The range estimate assumes 1\% of total
mass is radiated as gravitational waves, in rough agreement with
numerical simulations. Unlike neutron star inspirals, the abundance of
such ``intermediate mass black holes'' and hence their merger rate is
difficult to predict \cite{Flanagan:1997sx}. 

Searches are also in progress for GWs from CBCs with total masses up
to 100 solar masses, and from CBCs coincident with short-hard GRBs
observed during the S5 run. In addition, procedures are being
developed for establishing confidence in candidate detection events,
and for extracting the physical parameters of detected events.


\subsection{Gravitational-wave Bursts}

In addition to the well-modelled signals described in previous
sections, we search for gravitational-wave ``bursts'', defined as any
short-duration signal ($t \lesssim 1$~s) with significant signal power
in the detectors' sensitive frequency band ($45 \leq f \leq
2,000~$Hz). For example, the collapsing core of a massive star (the
engine that powers a type II supernova) can emit GWs through a number
of different mechanisms~\cite{Ott:2008wt}.  A compact binary
merger\,--\,discussed in the earlier section about CBC
searches\,--\,may be considered a burst, especially if the mass is
large so that the bulk of the long inspiral signal is below the
sensitive frequency band of the detectors, leaving only a short signal
from the actual merger to be detected.  Cosmic strings, if they exist,
are generically expected to bend into cusps and kinks which are
efficient radiators of beamed GWs. There may well be other
astrophysical sources, since any energetic event that involves an
asymmetric reshaping or re-orientation of a significant amount of mass
will generate GWs.

Many energetic gravitational events will also emit electromagnetic
radiation and/or energetic particles that can be observed with
telescopes and other astronomical instruments, as in the case of
supernovae.  Thus, besides searching for GW signals alone, we can
search for a class of joint emitters and use information from
conventional observations to constrain the GW event time and sky
position, allowing a more sensitive ``externally triggered'' search.
For example, Gamma-ray bursts (GRBs), and soft gamma-ray repeater
(SGR) flares are highly energetic events that make excellent targets for
externally-triggered GW burst searches.  While the progenitor(s) of
GRBs are not entirely clear, most if not all short-duration GRBs are
thought be produced by mergers of neutron stars or of a neutron star
with a black hole, which would radiate a great deal of energy in
GWs. Similarly, SGRs are believed to be neutron stars with very high
magnetic fields ({\it i.e.}\ magnetars) that sporadically produce
flares of electromagnetic radiation. The flares may be related to
deformations of the neutron star crust which could couple to GW
emission.  If an associated GW signal for these progenitors is
detected, the combined GW and EM/particle data will reveal
complementary information about the astrophysics of the event.

\subsubsection{Analysis methods}

A number of robust burst detection methods have been developed that
do not rely on knowledge of the signal waveform. Most fit into one of
three general categories: excess power, cross-correlation, or
coherent.

\textit{Excess power} methods decompose the data into different
frequency components, either with a Fourier basis or with some family
of wavelets, and look for signal power that is significantly above the
baseline detector noise level in some time-frequency region.  An excess
power method typically generates triggers from each detector, and then
applies a coincidence test to find consistent event candidates with excess power in
two or more detectors.

\textit{Cross-correlation} methods directly compare the data streams
from a pair of detectors to look for a common signal within
uncorrelated noise.  A cross-correlation statistic is calculated by
integrating over a short time window\,--\,ideally, comparable in length
to the duration of the signal\,--\,with a range of relative time delays
corresponding to different GW arrival directions.  The
cross-correlation is insensitive to the relative amplitude of the
common signal in the two data streams which may be different due to
the antenna response of the detectors.

\textit{Coherent} methods generalize the concepts of excess power and
cross-correlation to take full advantage of having three or more data
streams.  Detectors at each site see a different linear combination of
the same two time-dependent GW polarization components, so a network
of detectors at three sites (e.g. the two LIGO sites plus Virgo) has
enough information to over-determine the waveform and provide a
consistency test for each hypothetical arrival direction.  This is
essentially a maximum likelihood approach on the space of possible GW
signals, except that a ``regulator'' or Bayesian prior is used to
disfavor physically unlikely scenarios \cite{0264-9381-23-19-S05,
  0264-9381-25-10-105024}.  If only two sites are available, the use
of this regulator allows a somewhat weaker coherent analysis to be
performed on data from only two detectors.  In externally triggered
searches, coherent analysis is simpler because the sky location of the
potential signal is already known.  In this case two sites are
sufficient to fully determine the GW signal.

Each of these analysis methods produces a statistic (or more than one)
that describes the ``strength'' of the event candidate.  The strength
statistics are compared to the background distribution using time
shift analysis, like the CBC searches. Externally triggered
searches also determine the background from time shifted
off-source data.

These  search methods generally work well for a wide range
of signals, with some waveform-dependent variation between methods.
They are less sensitive than matched filtering for a known signal
but are computationally efficient and are often within a factor of 2
in sensitivity.

\subsubsection{Analysis results: All-sky burst searches}

The most general searches are those that look for GW bursts coming
from any sky position at any time.  Because there is no morphological
distinction between a GW burst signal and an instrumental glitch,
these ``all-sky'' searches place stringent demands on data quality
evaluation, instrumental veto conditions, and consistency tests among
detectors.  The primary S4 all-sky burst
search~\cite{0264-9381-24-22-002} was designed to detect signals with
frequency content in the range 64\,--\,1600~Hz and durations of up to
$\sim\!1$~s.  It identified event candidates with an analysis pipeline
consisting of two stages.  First, a wavelet-based excess power method
was used to find instances of coincident excess power in all three
detectors. Second, candidate triggers were generated with highly
significant correlation compared to background as well as positive
definite correlation and strain amplitude between the two Hanford
detectors.  No significant event candidates were identified in
15.53~days of observation; based on this, we placed an upper limit at
90\% confidence on the rate of detectable GW bursts of $0.15$ per day.

\begin{figure}[bt]
\begin{center}
\includegraphics[width=0.90\linewidth]{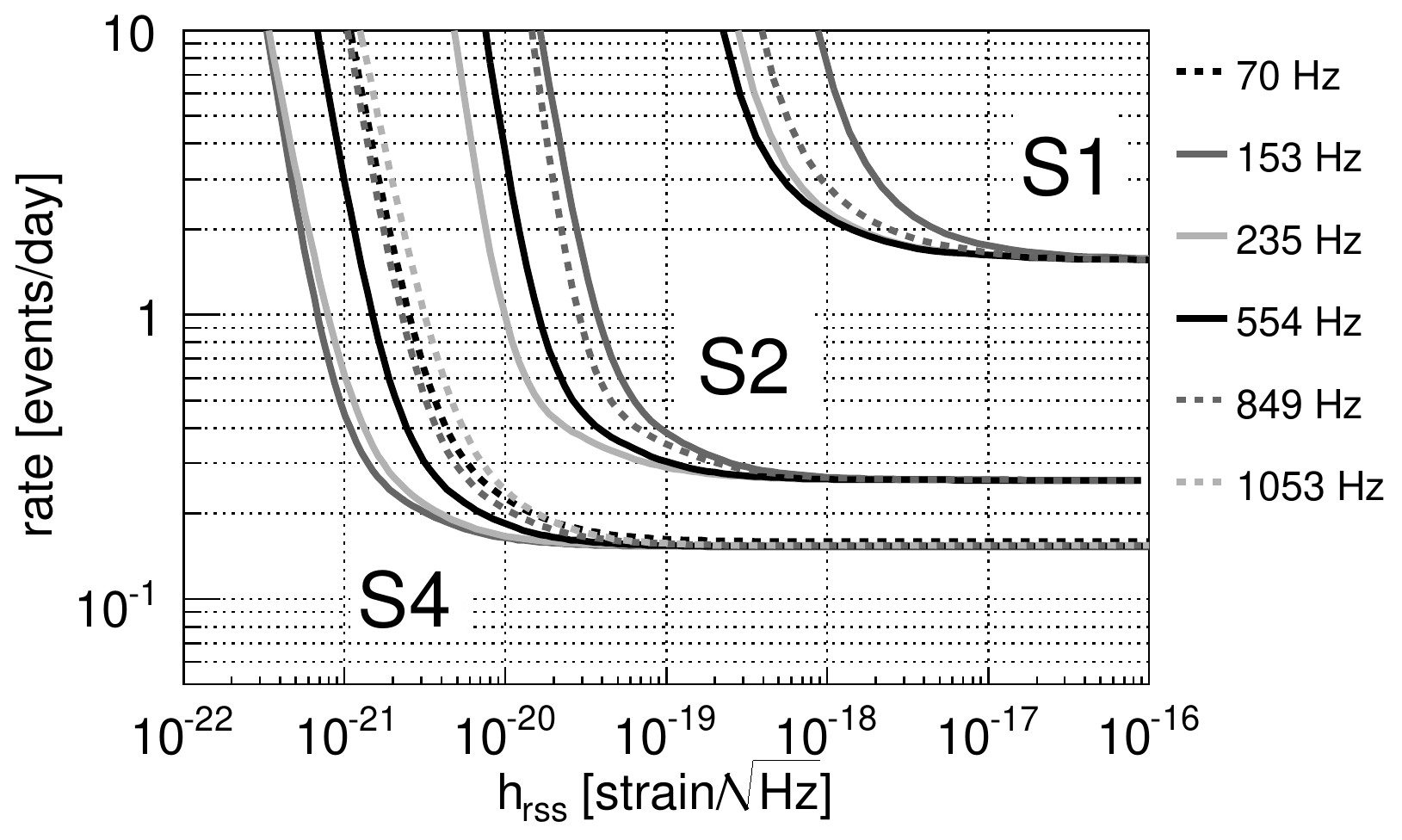}
\caption{Exclusion diagram (rate limit at 90\% confidence level, as a
function of signal amplitude) for sine-Gaussian signals with $Q=8.9$.
Search results from the S1, S2 and S4 science runs are shown.  (A
burst search was also performed for S3, but it used only 8 days of
data and systematic studies were not not carried through to produce a
definitive rate limit.)}
\label{fig:S1S2S4excl}
\end{center}
\end{figure}

To interpret a null result such as this one, a Monte Carlo method
evaluates what signals could have been detected by the search.  The
data are re-processed with simulated GW signals using the same analysis
pipeline to measure the detection efficiency in the presence of actual
detector noise. The intrinsic amplitude of a simulated burst signal is
characterized by a model-independent quantity, the ``root-sum-square''
GW strain, $h_\mathrm{rss}$, that expresses the amplitude of the GW
signal arriving at the Earth without regard to the response of any
particular detector.  It has units of $\mathrm{Hz}^{-1/2}$, allowing
it to be directly related to the amplitude spectral density of the
detector noise as shown in Fig.~\ref{fig:S1S2S4excl}.

In principle, the efficiency of a burst search pipeline can be
evaluated for any modeled GW waveform, {\it e.g.}\ from a core
collapse simulation or a binary merger signal generated using
numerical relativity.  In practice, the search efficiency is evaluated
for a collection of {\it ad hoc} waveforms that have certain general
features but do not correspond to any particular physical model. One
of our standard waveforms is a ``sine-Gaussian'', a sinusoidal signal
with central frequency $f$ within a Gaussian envelope with
dimensionless width parameter $Q$.
%
%
Evaluating the detection efficiency as a function of frequency,
Fig.~\ref{fig:S1S2S4excl} shows the effective rate limit as a
function of signal strength using an ``exclusion diagram''.

To understand the reach of the analysis in astrophysical terms, the
search sensitivity in terms of $h_\mathrm{rss}$ can be related to a
corresponding energy emitted in gravitational waves, $E_{\rm GW}$.  As
discussed in the S4 all-sky burst search
paper~\cite{0264-9381-24-22-002}, for sine-Gaussians and other
quasiperiodic signals,
\begin{equation}\label{eq:SGenergy}
E_\mathrm{GW} \sim \frac{r^2 c^3}{4 G} (2\pi f)^2 h_\mathrm{rss}^2
\end{equation}
where the GW energy emission is assumed to be isotropic.  GW emission
is not isotropic, but the energy flux varies by a factor of no more
than 4.  Using the fact that the S5 data has lower noise than S4 by
approximately a factor of two, sources at a typical Galactic distance
of 10~kpc could be detected for energy emission in GWs as low as $\sim
5 \times 10^{-8}~M_\odot$.  For a source in the Virgo galaxy cluster,
approximately 16~Mpc away, GW energy emission as low as $\sim
0.12~M_\odot$ could be detected.

We can draw more specific conclusions about detectability for models
of astrophysical sources that predict the absolute energy and
waveform emitted.  Following the discussion
in~\cite{0264-9381-24-22-002}, we estimate that a similar burst search
using S5 data could detect the core-collapse signals modeled by Ott
{\it et al.}~\cite{Ott:2006qp} out to $0.4$~kpc for their 11~$M_\odot$
non-spinning progenitor (model s11WW) and to $16$~kpc for their
15~$M_\odot$ spinning progenitor (model m15b6).  The
latter of these would be detectable throughout most of our Galaxy.  A
merger of two 10~$M_\odot$ black holes would be detectable out to a
distance of approximately $3$~Mpc, while a merger of two 50~$M_\odot$
black holes could be detected as far away as $\sim 120$~Mpc.

\subsubsection{Analysis results: Externally-triggered burst searches}

The exceptionally intense GRB 070201 was a particularly interesting
event for a triggered burst search because the sky position,
determined from the gamma-ray data, overlapped one of the spiral arms
of the large, nearby galaxy M31 (Andromeda).  An analysis of GW
data~\cite{Abbott:2007rh} found no evidence of an inspiral or a more
general burst signal; that finding ruled out (at the $\sim 99$\%
level) the possibility of a binary merger in M31 being the origin of
GRB 070201.

  We have searched for GW bursts associated with the giant
flare of SGR 1806$-$20 that occurred on December 27, 2004 (between the
S4 and S5 runs, but at a time when the LIGO H1 detector was operating,
albeit with reduced sensitivity) plus 190 smaller flares of SGRs
1806$-$20 and 1900$+$14 during the S5 run~\cite{Abbott:2008gj}.  No GW
signals were identified.  Energy emission limits were established
for a variety of hypothetical waveforms, many of them within the
energy range allowed by some models, and some as low as $3 \times
10^{38}$ J.  Future observations\,--\,especially for giant flares,
and flares of the recently-discovered SGR 0501$+$4516 which is closer
to Earth\,--\,will be sensitive to GW energy emission at or below the
level of observed electromagnetic energies.

Externally triggered GW burst searches are in progress or planned
using observations of supernovae, anomalous optical transients, radio
bursts, and neutrinos as triggers.  In general, the constraints on
event time, sky position, and (possibly) signal properties provided by
the external triggers make these searches a few times more sensitive
in amplitude than all-sky searches.  It is thus possible to
investigate a rich population of energetic transient events that may
plausibly produce detectable gravitational waves.

\subsection{Continuous Wave Sources}
Continuous GW signals may be generated by rotating neutron star such
as those powering millisecond radio pulsars.  In these systems, a
quadrupole mass asymmetry, or ellipticity, $\epsilon$, radiates GWs at
twice the neutron star rotation frequency.  The maximum sustainable
ellipticity, and hence the maximum GW emission, is a function of the
neutron star's internal structure and equation of state.  Current
limits on the ellipticity are based on the change in frequency of the
radio pulsar signal, the spin down rate, assuming that the spin down
is entirely due to GW emission.  An especially interesting candidate
is the Crab pulsar, for which the spin-down bound on ellipticity is
$\epsilon \leq 7.2 \times 10^{-4}$ and for which the bound on
detectable strain is $h \leq 1.4\times10^{-24}$ at about 59.6~Hz,
twice its spin frequency \cite{Abbott:2008fx}. ``Standard'' neutron
star equations of state predict $\epsilon \leq 10^{-7}$, while exotic
pulsars such as quark stars may have $\epsilon \leq 10^{-4}$
\cite{Owen:2005fn}.  For most radio pulsars, however, the spin down
limit overestimates the ellipticity and associated GW emission because
of electromagnetic damping of the rotation.

Compared with CBCs or bursts, neutron star powered millisecond radio
pulsars are a weak source of GWs which LIGO can detect only if the
source is within a few hundred parsecs.  Nonetheless, there are dozens
of known sources within this range that may be detected if they have
sufficiently high ellipticity. Furthermore, millisecond pulsars are
attractive sources of continuous GWs since the stable rotation periods
allow coherent integration over many hours, weeks and months to
improve the signal to noise ratio.

\subsubsection{Analysis methods}
\label{sec:analysis-methods-pulsar}

The shape of a rotating neutron star's detected GW waveform is a
function of at least six source parameters: two each for the pulsar
position and orientation on the sky, and at least two for the spin
frequency and frequency drift (1st time derivative). 
The intrinsic phase of a spinning neutron star waveform as measured in
the neutron star's rest frame, $\Phi(T)$, is modelled as an
approximate sinusoid at instantaneous frequency $\nu$ and spin-down
rate $\dot{\nu}$.  The observed phase in the detector frame, $\phi(t)$,
is in general a more complicated function of time due to the variable
time delay $\delta t = T - t$. The delay $\delta t$ contains
components arising from the Earth's orbital motion (for which $|\delta
t|\leq 8.5$~min.), from the Earth's sidereal motion ($|\delta t| \leq
43\,\mu$s), and from the general relativistic Shapiro delay ($|\delta
t| \leq 120\,\mu$s) for signals passing close to the
sun~\cite{Backer:1986wa}.

The six-dimensional parameter space and long duration of the S5 run
makes all-sky coherent searches for unknown neutron stars, for which
the amplitude and phase variations are tracked throughout the
observation time, computationally prohibitive.  Three techniques that
trade off between sensitivity and computation have been implemented:
1) semi-coherent, long duration all-sky searches sensitive only to
power and neglecting phase using the entire data set
\cite{Abbott:2007tda}; 2) coherent, short-duration all-sky searches
sensitive to amplitude and phase but computationally limited to
$\approx 5000$~hr integration time \cite{Abbott:2006vg}; 3) coherent,
targeted searches for millisecond radio pulsars with accurate and
stable ephemerides using the entire data set \cite{Abbott:2007ce}.
Although the coherent targeted search is the most sensitive, only a
little more than 100 known radio pulsars have suitable ephemerides,
while neutron star formation rates predict many hundreds of
millisecond pulsars within a detectable volume
\cite{Collaboration:2008rg}. Thus even though the all-sky searches are
not as sensitive as the targeted search, they are worth performing.

The LIGO Scientific Collaboration has implemented several different
incoherent all-sky searches. The most recent results using the S5 data
are from PowerFlux \cite{Collaboration:2008rg}.  The search
averages strain power from short Fourier transforms (SFTs) over the
full run to look for excess power in a narrow frequency bin.  The SFTs
are calculated using contiguous 30-minute data segments.  Before
summing, each SFT is shifted by a sky position dependent factor to
account for the time delays discussed above, and weighted according to
the detector antenna response and average noise power.  Frequency bins
with high SNR are checked for coincidence between multiple detectors
and followed up with coherent searches.

An alternative all-sky search using longer coherence times ($>$1 day)
offers improved sensitivity, but its computational demands require a
new paradigm: distributed computing using the Einstein@Home network
\cite{EinsteinWebpage}.  Einstein@Home users volunteer their idle
computing CPU cycles to perform a coherent analysis.  The combined
resources of 50,000 volunteers with 100,000 CPUs enables an all-sky
search for rotating neutron stars in 5280 hours of the most sensitive
S5 data.  The Einstein@Home search is based on the coherent
$\mathcal{F}$-statistic in the frequency
domain~\cite{Jaranowski:1998qm}.  Each CPU in the distributed network
calculates the coherent signal for each frequency bin and sky position
for a 30-hour contiguous segment. The loudest frequency bins are
followed up with coincidence studies between detectors and continuity
studies with adjacent time segments.

The deepest searches are performed for millisecond radio pulsars with
well-characterized, stable ephemerides.  The 154 pulsars with spin
frequencies greater than 25 Hz are selected from the Australian
Telescope National Facility online catalogue \cite{ATNFcatalog}.  Of
these, 78 have sufficient stability and timing resolution to make
knowledge of their waveform improve the detection SNR over the long
observation time.  To consistently incorporate the prior information,
the targeted search uses a time-domain, Bayesian analysis in which the
detection likelihood is calculated for each detector.  Information
from multiple detectors is combined to form a joint likelihood
assuming the detectors' noises are independent.  This procedure allows
upper limits from successive science runs to be combined and provides
a natural framework for incorporating uncertainties in the
ephemerides.

\subsubsection{Analysis results}
\label{sec:analysis-results-pulsar}

Analyses of the full S5 data are underway using the techniques
described above. An all-sky search using the PowerFlux technique on
the first 8 months of S5 with the H1 and L1 detectors has been
completed~\cite{Collaboration:2008rg}. This produced upper limits on
strain amplitude in the band 50\,--\,1100~Hz.  For a neutron star with
equatorial ellipticity of $10^{-6}$, the search was sensitive to
distances as great as 500 pc.

\begin{figure}[hbt]
  \centering
  \includegraphics[width=4in]{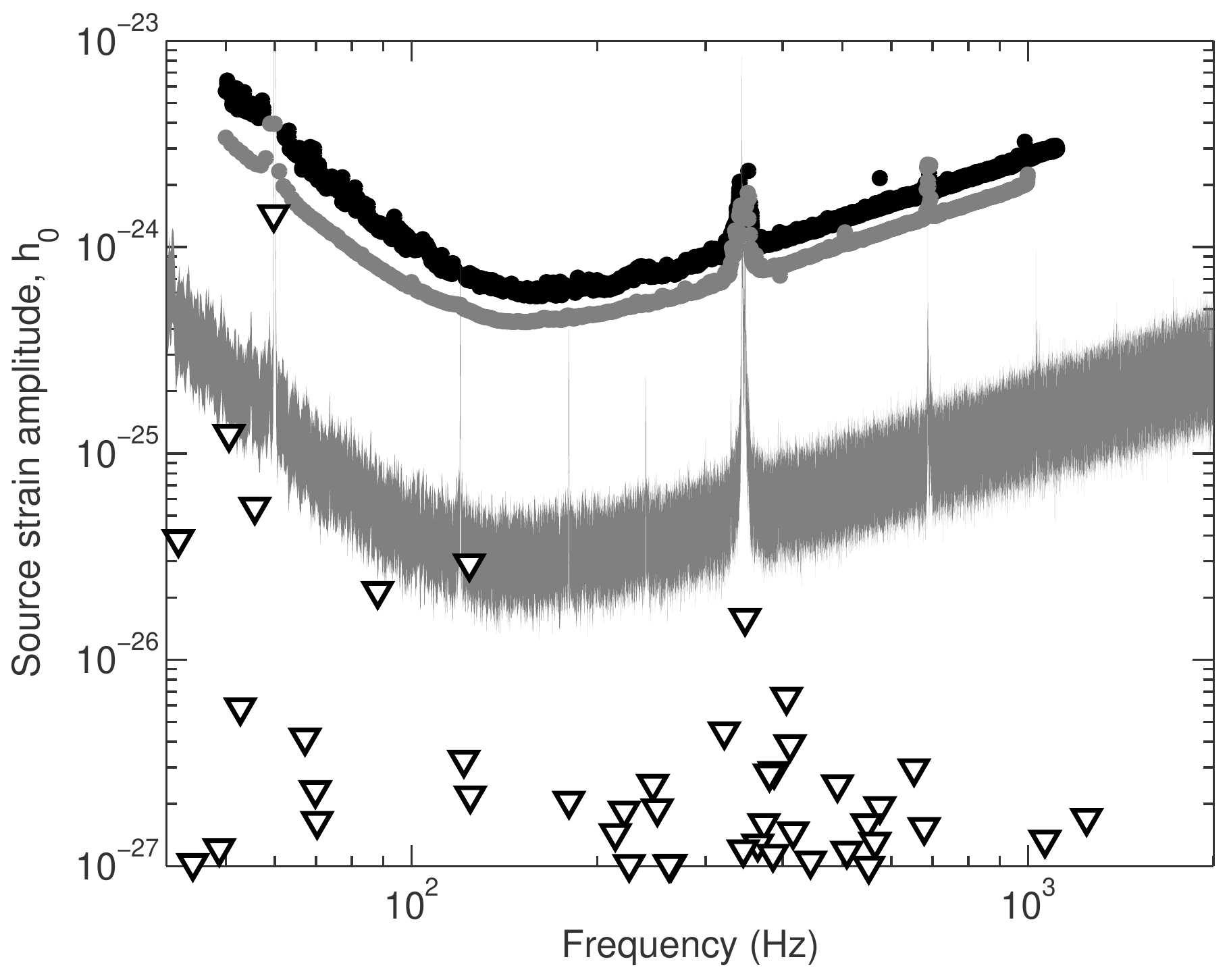}
  \caption{Limits on GW strain from rotating neutron stars. Upper
    curve (black points): all-sky strain upper limits on unknown
    neutron stars for spindown rates as high as $5\times10^{-9}$ Hz
    s$^{-1}$ and optimal orientation, from analysis of the first 8
    months of S5 data~\cite{Collaboration:2008rg}. Middle curve (gray
    points): expected sensitivity for the Einstein@Home search with
    5280 hrs of S5 data. Lower curve (gray band): expected range for
    95\% confidence level Bayesian upper limits on radio pulsars with
    known epherimides, using the full S5 data. Black triangles: upper
    limits on GW emission from known radio pulsars based on their
    observed spin-down rates.  }
  \label{fig:pulsarlimits}
\end{figure}

Because of the narrow bandwidth ($10^{-6}$~Hz) and complicated
frequency modulation of pulsar signals, instrument artifacts do not
significantly contribute to the noise in pulsar searches.  The few
exceptions\,--\,non-stationary noise near 60~Hz harmonics, wandering
lines, etc.\,--\,have been easy to identify and remove.  Consequently
we can predict the astrophysical reach of the full S5 data set with a
high degree of confidence based on the performance of previous
searches and the S5 noise performance. Figure~\ref{fig:pulsarlimits}
shows the projected S5 strain amplitude sensitivity for the more
sensitive searches, along with the upper limits established by the
PowerFlux analysis.


\subsection{Stochastic GW Background}
A stochastic background of gravitational waves could result from the
random superposition of an extremely large number of unresolved and
independent GW emission events \cite{1997rggr.conf..373A}. Such a
background is analogous to the cosmic microwave background radiation,
though its spectrum is unlikely to be thermal. The emission events
could be the result of cosmological phenomena, such as the
amplification of vacuum fluctuations during inflation or in
pre-big-bang models; phase transitions in the history of the universe;
or cosmic strings, topological defects that may have been formed
during symmetry-breaking phase transitions in the early universe. Or a
detectable background could result from many unresolved astrophysical
sources, such as rotating neutron stars, supernovae, or low-mass X-ray
binaries.

Theoretical models of such sources are distinguished by the power
spectra they predict for the stochastic background production. The
spectrum is usually described by the dimensionless quantity
$\Omega_{\rm GW}(f)$, which is the GW energy density per unit
logarithmic frequency, divided by the critical energy density $\rho_c$
to close the universe:
\begin{eqnarray}
  \Omega_{\rm GW}(f) = \frac{f}{\rho_c}\frac{d \rho_{\rm GW}}{df} \,.
\end{eqnarray}
In the LIGO frequency band, most of the model spectra are well
approximated by a power-law: $ \Omega_{\rm GW}(f) \propto f^\alpha$.
LIGO analyses consider a range of values for $\alpha$, though in this
review we will focus on results for a frequency independent
$\Omega_{\rm GW}$ ($\alpha=0$), since many of the cosmological models
predict a flat or nearly flat spectrum over the LIGO band.  The strain
noise power spectrum for a flat $\Omega_{\rm GW}$ falls as $f^{-3}$,
with a strain amplitude scale of\footnote[6]{We assume here and in the
  rest of this paper a Hubble expansion rate of 72~km/sec/Mpc.}:
\begin{eqnarray}
  h_{\rm GW} = 4 \times 10^{-22} \sqrt{\Omega_{\rm GW}}\, 
        \left(\frac{{\rm 100\,Hz}}{f}\right)^{3/2} {\rm Hz}^{-1/2} \,.
\end{eqnarray}

\subsubsection{Analysis method}
\label{sec:analysis-methods-stoch}

Unlike the cosmic microwave background, any GW stochastic background
will be well below the noise floor of a single detector. To probe
below this level, we cross-correlate the output of two detectors
\cite{Allen:1997ad}.  Assuming the detector noises are independent of
each other, in the cross-correlation measurement the
signal\,--\,due to a stochastic background present in each
output\,--\,will increase linearly with integration time $T$, whereas
the measurement noise will increase only with the square root of
$T$. Similarly the signal will increase linearily with the effective
bandwidth ($\Delta f$) of the correlation, and the noise as $(\Delta
f)^{1/2}$. Thus with a sufficiently long observation time and wide
bandwidth, a small background signal can in principle be detected
beneath the detector noise floor.

The assumption of independent detector noise is crucial, and it is 
valid when comparing L1 with either of the Hanford detectors due to
their wide physical separaion. But this separation also extracts a
price: the coherent cross-correlation of a stochastic
background signal is reduced by the separation time delay between the
detectors and the non-parallel alignment of their arms. These effects
are accounted for by the \textit{overlap reduction function}
$\gamma(f)$, which is unity for co-located and co-aligned detectors,
and decreases below unity when they are shifted apart or mis-aligned.
For a Livingston-Hanford detector pair, the overlap is on average
$\langle\gamma\rangle \sim 0.1$ in the sensitive band around 100~Hz.

The low frequency noise floor of a single S5 LIGO detector is
roughly equivalent to a stochastic background spectrum with $\Omega_{\rm GW} =
0.01$ ($h_{\rm GW} = 4 \times 10^{-23}\,{\rm Hz}^{-1/2}$ at 100~Hz).
The cross-correlation measurement will be
sensitive to a background $\Omega_{\rm GW}$ lower than this noise
floor by a factor of $\langle\gamma\rangle (T \Delta
f)^{1/2}$. With a year of observation time and an effective bandwidth
of 100~Hz this is a factor of several thousand, so we expect to probe
for a stochastic background in the range $\Omega_{\rm GW} \sim 10^{-5}
- 10^{-6}$.

\subsubsection{Analysis results}

Since the SNR for a search on $\Omega_{\rm GW}$ grows inversely with
the product of the strain noise amplitude spectra of the two
detectors, the sensitivity of this search grew quickly as the
detectors improved. Analysis of the S4 data yielded a Bayesian 90\%
upper limit of $\Omega_{\rm GW} < 6.5 \times 10^{-5}$ for a flat
spectrum in the 51\,--\,150~Hz band \cite{Abbott:2007sgwb}.  Projecting
for the S5 data, the lower strain noise and longer integration time
should improve on this by an order of magnitude. While the
cross-correlation analysis for S5 is still in progress, it is
straightforward to calculate the expected variance of the
cross-correlation using only the interferometers' strain noise spectra
over the run.  This predicts that the potential upper limit on
$\Omega_{\rm GW}$ will be in the range $4-5 \times 10^{-6}$.

Such a result would be the first direct measurement to place a limit
on $\Omega_{\rm GW}$ more stringent than the indirect bound set by
Big-Bang-Nucleosynthesis (BBN). The BBN bound, currently the most
constraining bound in the band around 100~Hz, derives from the fact
that a large GW energy density present at the time of BBN would have
altered the abundances of the light nuclei in the universe
\cite{1997rggr.conf..373A, Maggiore:1999vm}. For the BBN model
to be consistent with observations of these abundances, the total GW
energy density at the time of nucleosynthesis is thus constrained. In
the limiting case that the GW energy was confined to LIGO's sensitive
band of 50\,--\,150~Hz, the BBN bound is: $\Omega_{\rm GW} < 1.1 \times
10^{-5}$ \cite{Maggiore:1999vm, Cyburt2005313}.

\begin{figure}[hbt]
  \centering
  \includegraphics[width=5in]{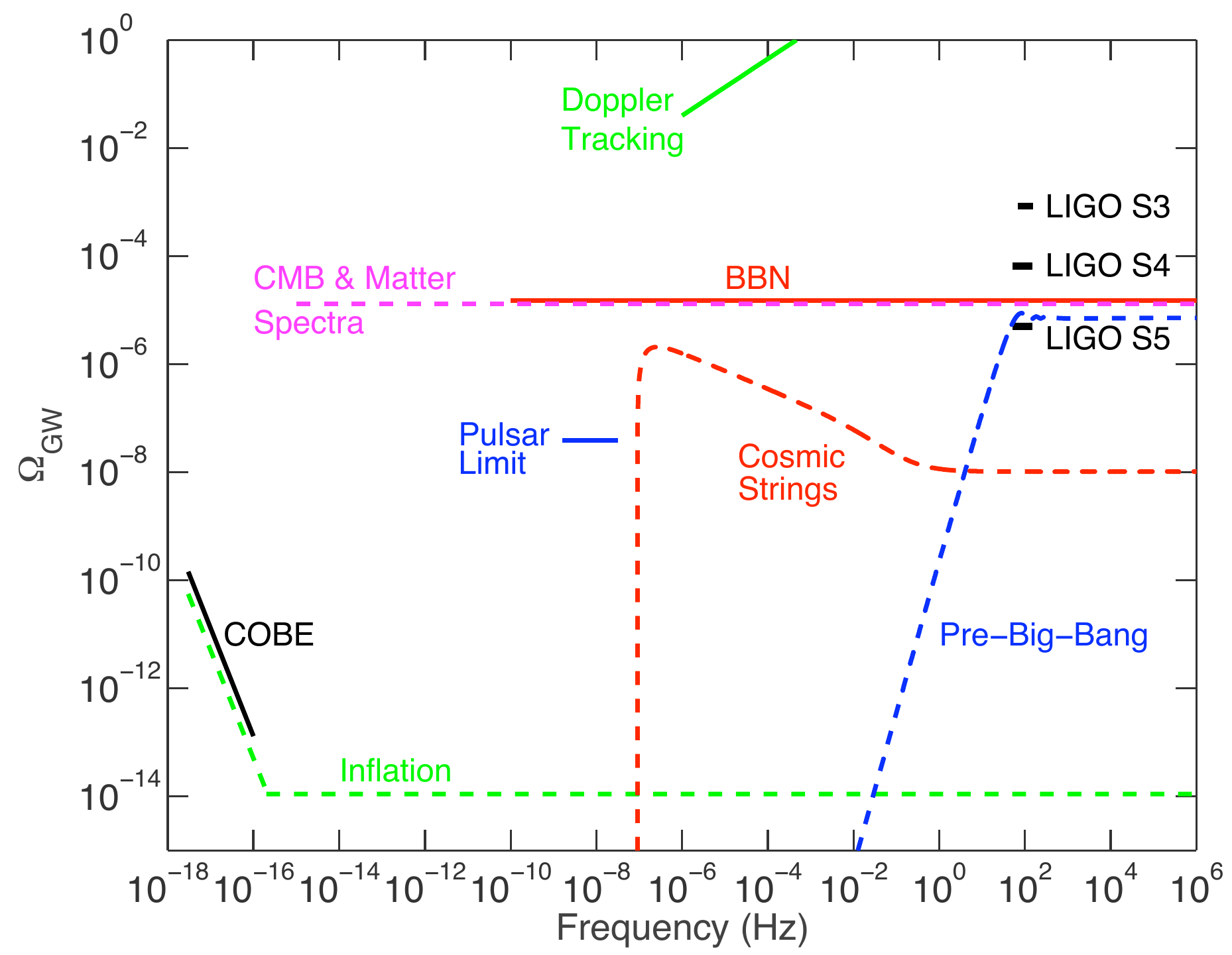}
  \caption{Observational limits and potential sources for a stochastic 
   background of gravitational waves. The LIGO S5 curve refers to the
   potential upper limit from the S5 run, based on strain noise data. 
   The curves corresponding to
   inflationary, cosmic-string and pre-big-bang models are examples;
   the model parameters allow significant variation in the predicted
   spectra. The BBN and CMB \& Matter Spectra bounds apply to the
   total GW energy over the frequency range spanned by the 
   corresponding lines. See reference \cite{Abbott:2007sgwb} for
   more details.}
  \label{fig:Landscape}
\end{figure}

The LIGO results are also being used to constrain the
parameter space of models predicting a stochastic GW background, such as
cosmic string models and pre-big-bang models \cite{Abbott:2007sgwb}.
The gamut of theoretical models and observations pertaining to a
stochastic GW background spans an impressively wide range of
frequencies and amplitudes. These are displayed in the landscape plot
of Fig.~\ref{fig:Landscape}.

The analysis described so far searches for an isotropic background of GWs. 
The cross-correlation method has also been extended to search for spatial
anisotropies, as might be produced by an ensemble of astrophysical sources 
\cite{Ballmer:rad}. Such a GW radiometer requires spatially separated
interferometers in order to point the multi-detector antenna at different
locations on the sky. The result is a map of the power distribution of
GWs convolved with the antenna lobe of the radiometer, with an uncertainty
determined by the detector noise. Radiometer analysis of the S4 data 
yielded upper limits on the GW strain power from point sources in the
range of $\sim\!10^{-48}\,{\rm Hz}^{-1}$ to $\sim\!10^{-47}\,{\rm Hz}^{-1}$,
depending on sky position and the GW power spectrum model \cite{Abbott:082003}.
The S5 analysis should improve on the strain power sensitivity by a
factor of 30. The corresponding GW energy flux density that the
search will be sensitive to is 
$\sim\!10^{-10}\,{\rm Watt/m^2/Hz}\,(f/{\rm 100\,Hz})^2$.



\section{Future directions}  
From its inception, LIGO was envisioned not as a single experiment,
but as an on-going observatory. The facilities and infrastructure
construction were specified, as much as possible, to accommodate
detectors with much higher sensitivity.  We have identified a set of
relatively minor improvements to the current instruments
\cite{LIGO:T060156} that can yield a factor of 2 increase in strain
sensitivity and a corresponding factor of 8 increase in the probed
volume of the universe. The two most significant enhancements are
higher laser power and a new, more efficient readout technique for the
GW channel. Higher power is delivered by a new master oscillator-power
amplifier system, emitting 35~W of single frequency 1064~nm light
\cite{Frede:07}, 3.5 times more power than the initial LIGO
lasers. For the readout, a small mode-cleaner cavity is inserted in
the AS beam path, between the Faraday isolator and the length
photodetectors. This cavity filters out RF sidebands and the
higher-order mode content of the AS port light, reducing the
shot-noise power. Instead of RF heterodyning, signal detection is done
by slightly offsetting the differential arm length from the dark
fringe, and using the resulting carrier field as the local oscillator
in a DC homodyne detection scheme.  These improvements (known
collectively as Enhanced LIGO) are currently being implemented and
commissioned on H1 and L1, and a one-to-two year science run with
these interferometers is expected to begin in mid-2009.

Significantly greater sensitivity improvements are possible with more
extensive upgrades. Advanced LIGO will replace the exisiting
interferometers with significantly improved technology to achieve a
factor of at least 10 in sensitivity over the initial LIGO
interferometers and to lower the seismic wall frequency down to 10~Hz
\cite{2003SPIE.4856..282F, LIGO:M060056}.  Advanced LIGO has
been funded by the National Science Foundation, begining in April
2008. Installation of the Advanced LIGO interferometers is planned to
start in early-2011.

The Advanced LIGO interferometers are configured like initial
LIGO\,--\,a power-recycled Fabry-Perot Michelson\,--\,with the
addition of a \textit{signal recycling} mirror at the anti-symmetric
output. Signal recycling gives the ability to tune the interferometer
frequency response, so that the point of maximum response can be
shifted away from zero frequency \cite{Meers:1988wp}.  The laser
wavelength stays at 1064~nm, but an additional high-power stage brings
the laser power up to 200~W \cite{2008CQGra..25k4040W}. The test
masses will be significantly larger\,--\,40~kg\,--\,in order to reduce
radiation pressure noise and to allow larger beam sizes.  Larger beams
and better dielectric mirror coatings combine to reduce the test mass
thermal noise by a factor of 5 compared to initial LIGO 
\cite{0264-9381-24-2-008}.

The test mass suspensions become significantly more intricate to
provide much better performance. They incorporate four cascaded stages
of passive isolation, instead of just one, including vertical
isolation comparable to the horizontal isolation at all stages except
one \cite{2002CQGra..19.4043R}.  The test mass is suspended at the
final stage with fused silica fibers rather than steel wires; these
fibers have extremely low mechanical loss and will reduce suspension
thermal noise nearly a hundred-fold \cite{2006PhLA..354..353H}. The
current passive seismic isolation stacks that support the suspensions
are replaced with two-stage active isolation platforms
\cite{0264-9381-19-7-349}. These stages are designed to actively
reduce the ground vibration by a factor of $\sim\!1000$ in the
$1-10$~Hz band, and provide passive isolation at higher
frequencies. The combination of the isolation platforms and the
suspensions will reduce seismic noise to negligible levels above
approximately 10~Hz.

The successful operation of Advanced LIGO is expected to transform the 
field from GW detection to GW astrophysics.  We illustrate the potential
using compact binary coalescences. Detection rate estimates for CBCs 
can be made using a combination of extrapolations from observed binary 
pulsars, stellar birth rate estimates, and population synthesis models. 
There are large uncertainties inherent in all of these methods, however,
leading to rate estimates that are uncertain by several orders of magnitude.
We therefore quote a range of rates, spanning plausible pessimistic and 
optimistic estimates, as well as a likely rate. For a NS mass of $1.4 M_\odot$
and a BH mass of $10 M_\odot$, these rate estimates for Advanced LIGO are: 
$0.4-400\,yr^{-1}$, with a likely rate of $40\,yr^{-1}$ for NS-NS binaries; 
$0.2-300\,yr^{-1}$, with a likely rate of $10\,yr^{-1}$ for NS-BH binaries; 
$2-4000\,yr^{-1}$, with a likely rate of $30\,yr^{-1}$ for BH-BH binaries.

\ack
The authors gratefully acknowledge the support of the United States
National Science Foundation for the construction and operation of
the LIGO Laboratory and the Particle Physics and Astronomy Research
Council of the United Kingdom, the Max-Planck-Society and the State
of Niedersachsen/Germany for support of the construction and
operation of the GEO\,600 detector. The authors also gratefully
acknowledge the support of the research by these agencies and by the
Australian Research Council, the Natural Sciences and Engineering
Research Council of Canada, the Council of Scientific and Industrial
Research of India, the Department of Science and Technology of
India, the Spanish Ministerio de Educacion y Ciencia, The National
Aeronautics and Space Administration, the John Simon Guggenheim
Foundation, the Alexander von Humboldt Foundation, the Leverhulme
Trust, the David and Lucile Packard Foundation, the Research
Corporation, and the Alfred P. Sloan Foundation.

\bibliography{LIGO_RPP}
\bibliographystyle{unsrt}

\end{document}